\pdfoutput=1
\documentclass[a4paper,11pt]{article}

\usepackage{graphicx,multicol,subcaption,amsmath,amssymb,float,hyperref}
\usepackage[T1]{fontenc} 
\usepackage[section]{placeins}
\usepackage{url}
\usepackage{color}
\usepackage[compat=1.1.0]{tikz-feynman}
\usepackage[margin=1.0in]{geometry}
\bibliography{bibliography}
\usepackage{cleveref}
\usepackage{braket}
\usepackage{multirow}
\usepackage{array}
\usepackage{changepage}
\newcommand{\lagr}{\mathcal L}
\newcommand{\func}{\operatorname}

\newcommand{\Wm}{W_{\mu}}
\newcommand{\Zm}{Z_{\mu}}

\newcommand{\RE}{\operatorname{Re}}

\newcommand{\thQ}{\theta ^Q _{34}}

\begin{document}
	
\begin{titlepage}
	\begin{center}
		{\bf\Large
			\boldmath{
                    Discovering the Origin of Yukawa Couplings at the LHC \\ [0.20cm]
                    with a Singlet Higgs and Vector-like Quarks
			}
		} \\[12mm]
		Simon~J.~D.~King$^{\star}$\footnote{E-mail: \texttt{sjd.king@soton.ac.uk}},
		Stephen~F.~King$^{\star}$\footnote{E-mail: \texttt{king@soton.ac.uk}},
		Stefano~Moretti$^{\star}$\footnote{E-mail: \texttt{stefano@soton.ac.uk}},
		Samuel~J.~Rowley$^{\star}$\footnote{E-mail: \texttt{s.rowley@soton.ac.uk}},
		\\[-2mm]
	\end{center}
	\vspace*{0.50cm}
	\centerline{$^{\star}$ \it
		Department of Physics and Astronomy, University of Southampton,}
	\centerline{\it
		SO17 1BJ Southampton, United Kingdom }
	\vspace*{1.20cm}
	
	\begin{abstract}
	{\noindent
	Although the 125 GeV Higgs boson discovered at the LHC is often heralded as the origin of mass, it may not in fact be the origin of Yukawa couplings. In alternative models, Yukawa couplings may instead arise from a seesaw type mechanism involving the mixing of Standard Model (SM) chiral fermions with new vector-like fermions, controlled by the vacuum expectation value (VEV) of a new complex Higgs singlet field $\langle \Phi \rangle$. For example, the largest third family $(t,b)$ quark Yukawa couplings may be forbidden by a $U(1)'$ gauge or global symmetry, broken by $\langle \Phi \rangle$, and generated effectively via mixing with a vector-like fourth family quark doublet $(T,B)$.
	 Such theories predict a new physical Higgs singlet $\phi$, which we refer to as the Yukon,
	 resulting from $\langle \Phi \rangle$, in the same way that the Higgs boson $h^0$ results from $\langle H\rangle$. In a simplified model we discuss the prospects for discovering the Yukon $\phi$ in gluon-gluon fusion production, with $(t,b)$ and $(T,B)$ quarks in the loops, and decaying in the channels
		$\phi\rightarrow \gamma\gamma, Z\gamma$ and 
$\phi\rightarrow tT\rightarrow tth^0,ttZ$. 
		The potential for discovery of the Yukon $\phi$ is studied at present or future hadron colliders such as the LHC (Run 3), HL-LHC, HE-LHC and/or FCC. For example, we find that a 300-350 GeV Yukon $\phi$ could be accessed at LHC Run 3 in the di-photon channel in the global model, providing a smoking gun signature of the origin of Yukawa couplings. The $tth^0,ttZ$ channels are more involved and warrant a more sophisticated analysis.}
	\end{abstract}
\end{titlepage}

\section{Introduction}
Given the failure so far of the Large Hadron Collider (LHC), superseding the TeVatron machine by a tenfold increase in both energy and luminosity,  to detect any new particles Beyond the Standard Model (BSM), it is tempting to conclude that a so-called desert landscape awaits us in the search for new physics. In any case, to make progress will require substantial increase in machine energy and/or luminosity, such as considered for a High Energy LHC (HE-LHC) \cite{Dainese:2019rgk}, a High Luminosity LHC (HL-LHC) \cite{Dainese:2019rgk,Gianotti:2002xx} and/or a Future Circular Collider (FCC) \cite{Abada:2019lih}. 

However, the conclusion that the LHC will not discover new particles is certainly premature since there are certain kinds of BSM physics which are perfectly consistent with the SM and only constrained by the limits placed by direct collider searches. Indeed, such associated new particles could be lying in wait to be discovered already at Run 3 of the LHC. Such new physics, if it is to be not in conflict with current measurements, must have a certain property known as decoupling, such that the masses of the new heavy particles involved can be smoothly taken to be very large without spoiling the consistency of the SM precision measurements. 

A well-known example of decoupling theories is supersymmetry (SUSY) \cite{Chung:2003fi}, where the squarks and sleptons can be made arbitrarily heavy without affecting any such precision measurements. Another example, not as widely studied as SUSY, which concerns us here is that of theories with vector-like fermions, whose left- and right-handed components transform in the same representation of the SM gauge group.
Unlike a sequential chiral fourth family of quarks and leptons, which relies on the Higgs Yukawa couplings for its masses, and therefore cannot be made sufficiently heavy to have avoided collider constraints without violating some SM measurements (such as the SM Higgs production cross-sections and decay rates), a vector-like fourth family, having both left- and right-handed components transforming identically,
may be given large Dirac masses by hand. Furthermore, such masses may be increased at will without impacting on any other SM measurement.

The search for vector-like quarks (VLQs) is particularly interesting at the LHC~\cite{Aaboud:2018pii,Romao:2020ocr}
as they can readily be produced, being coloured fermions which are pair produced via their couplings to gluons, which are universal and identical to the standard QCD ones.
There are various SM assignments that the VLQs may take, the simplest case being that they have the SM assignments of the usual quarks, i.e. colour triplets being Electro-Weak (EW) doublets and singlets with the usual hypercharge assignments, resulting in the usual electric charges of the observed quarks. For example, the phenomenology of a VLQ EW doublet $(T,B)$ with the electric charges of the usual quark EW doublet $(t,b)$ has recently been studied in great detail in~\cite{Aguilar-Saavedra:2017giu}.
Indeed there are many motivations for VLQs of all kinds arising from theories of little Higgs \cite{Perelstein:2003wd,Schmaltz:2005ky}, composite Higgs \cite{Contino:2006qr,Contino:2006nn,Matsedonskyi:2012ym,Kaplan:1991dc},
SUSY \cite{Martin:2009bg,Martin:2010dc} and quark mixing ~\cite{Botella:2018aon,Botella:2016ibj}. 

Here we shall focus on a quite different motivation for VLQs with distinctive experimental implications, namely that they may play a crucial role in the origin of all SM Yukawa couplings, in such a way that all fermion masses would be zero in the absence of mixing with vector-like fermions. The first example  \cite{Ferretti:2006df} of such a model was one in which 
the usual Higgs Yukawa couplings with the three families of chiral fermions were forbidden
by a discrete $Z_2$ symmetry, which is subsequently broken by 
the vacuum expectation value (VEV) of a new complex Higgs singlet field $\langle \Phi \rangle$, allowing 
the Higgs Yukawa couplings to arise effectively from mixing with a vector-like fourth 
family. The origin of the large Yukawa couplings of the $(t,b)$ quarks, with small mixing,
was then explained by assuming that 
the mass of the VLQ EW doublet $(T,B)$ is much lighter than the EW singlet VLQs. 
The origin of the Yukawa couplings of the second family $(c,s)$ quarks is due to their mixing with other VLQs
which are EW singlets, whose masses are assumed to be much heavier than $(T,B)$.
From an LHC perspective, only the lightest VLQ EW doublet $(T,B)$ is relevant.

In a later example \cite{King:2018fcg}, the $Z_2$ symmetry above was replaced by a 
gauged $U(1)'$ symmetry, under which the SM fermions are neutral but the Higgs doublets are charged,
thereby forbidding Yukawa couplings but allowing mixing with the charged vector-like fourth family.
If the $U(1)'$ symmetry is gauged, its breaking will yield a massive $Z'$ gauge boson with non-universal couplings 
to quarks and leptons is generated by the mixing with the fourth family \cite{King:2018fcg}.
In such a model \cite{King:2018fcg}, the 
connection between non-universal $Z'$ couplings and the origin of Yukawa couplings may have interesting 
experimental implications for the $R_K$ and $R_{K^*}$ anomalies \cite{Aaij:2014ora}, since 
the lightest VLQ doublet $(T,B)$ will mix strongly with the $(t,b)$ quarks and generate
both Yukawa and $Z'$ couplings for the latter. 
In such models, the vector-like fourth family may emerge as a Kaluza-Klein excitation of quarks and leptons 
in 5d~\cite{deAnda:2020hcf}.
More recently a global $U(1)'$ version of such a model has been considered, focussing on the $g-2$ muon and electron anomalies in a two Higgs doublet model (2HDM) with fourth and fifth vector-like families~\cite{Hernandez:2021tii}.

In the above models, the presence of the new singlet Higgs $\Phi$ associated 
with the breaking of the extra symmetry ($Z_2$ or $U(1)'$), with  
coupling to both the SM quarks $(t,b)$ and VLQ doublet $(T,B)$,
may have interesting implications for collider studies which have not so far been considered in the literature.
In particular, such theories predict a new physical Higgs singlet $\phi$, resulting from $\langle \Phi \rangle$, in the same way that the Higgs boson $h^0$ results from $\langle H\rangle$.
The discovery of $\phi$ with the predicted couplings to VLQs would be tantamount to the discovery of the origin of Yukawa couplings, just as the discovery of the Higgs boson $h^0$ was said to be equivalent to the discovery of the origin of mass.  
For that reason we shall refer to Higgs singlet boson $\phi$ as the Yukawa boson or Yukon for short.
This motivates the present study in which we provide the first phenomenological analysis of the production and decay modes of the 
Yukon $\phi$ with the necessary couplings to VLQs $(T,B)$ in order to generate Yukawa couplings.
We shall see that what distinguishes this model from the SM plus VLQs is the existence of the Yukon singlet with couplings to fermions proportional to the fermion masses, a feature shared with that of the SM Higgs boson.

In the present paper, then, we consider the experimental signatures associated with the origin of Yukawa couplings along the lines
of the above models \cite{Ferretti:2006df,King:2018fcg,deAnda:2020hcf,Hernandez:2021tii}. 
We propose and study a simplified model in which 
we shall ignore all fermions apart from the third family $(t,b)$ quarks, since they mix most strongly with the lightest VLQ EW doublets $(T,B)$. In our simplified model
we suppose that the direct $(t,b)$ Yukawa couplings to Higgs doublets, in the limit of zero mixing with $(T,B)$,
 to be forbidden by either a gauge or a global $U(1)'$ symmetry broken by a Higgs singlet $\Phi$. The third family quark Yukawa
couplings are generated effectively after $U(1)'$ breaking via mixing with a vector-like fourth family quark doublet $(T,B)$. 
We shall focus on the resulting physical singlet Higgs Yukon $\phi$ associated with the $U(1)'$ breaking, with characteristic couplings
	to $(t,b)$ and $(T,B)$ quarks, whose discovery at hadron colliders would provide evidence for such models.
	We shall assume that there is negligible mixing of the $\Phi$ with the two Higgs doublets, so that the physical $\phi$ scalar boson predominantly arises as the real component of the complex singlet field $\Phi$ after it develops its VEV.
	This is a natural assumption in the case that $\langle \Phi \rangle $ greatly exceeds the Higgs doublet VEVs, and can be enforced by assuming certain coupling terms in the Higgs potential which couple $\Phi$ to the Higgs doublets to be small.
	We discuss the prospects for discovering the Yukon $\phi$
	at Run 3 of the LHC, HE-LHC and HL-LHC as well as a FCC, focussing in particular on 
		gluon-gluon Fusion (ggF) production, including both $(t,b)$ and $(T,B)$ quarks in the loops, of the Yukon $\phi$ which promptly decays via the channels
		$\phi\rightarrow \gamma\gamma, Z\gamma$ and 
$\phi\rightarrow tT\rightarrow tth^0,ttZ$\footnote{Hereafter, with our textual notation $tt$, $tT$, etc., we always intend the appropriate charge conjugated channels, i.e., $t\bar t$,  $t\bar T + \bar t T$, etc.}. 
		The discovery of the Yukon $\phi$ through any of the described production and decay modes at any of the mentioned colliders would shed new light on the origin of Yukawa couplings. 

Another interesting signature of the gauge model would be the $Z'$, which could be discovered in several similar channels to the Yukon $\phi$, such as $Z' \rightarrow tT \rightarrow tth^0,ttZ$. However the production will be suppressed as it can only be produced directly at the LHC through $b \bar{b} \rightarrow Z' $, or through ggF. 
In this paper, we will not pursue $Z'$ signatures but instead focus on the Yukon $\phi$, which would be a smoking gun signature of the origin of Yukawa couplings. It would be interesting to discuss the $Z'$ and 2HDM signatures in a future publication.

		It is worth briefly comparing our study to other studies of singlet Higgs production and decay at the LHC, related to flavon physics~\cite{Tsumura:2009yf,Berger:2014gga,Xiao:2014kba}. 
The study in~\cite{Tsumura:2009yf} focusses mainly on flavour changing processes and does not consider any vector-like quarks. Similarly the simplified models analysed in~\cite{Berger:2014gga,Xiao:2014kba} involve EW singlet ``top partners'' rather than the EW doublet vector-like quarks considered here. 
In such studies the Higgs singlet is called a ``flavon'' since it is related to the breaking of a flavour symmetry whose structure is responsible for the fermion mass hierarchies. We emphasise that in our case the $U(1)'$ symmetry is not a flavour symmetry since chiral quarks and leptons of all three families carry zero charges under it \cite{King:2018fcg}. Also, most realistic flavon models involve a large number of flavon fields, whereas in the type of model considered here \cite{Ferretti:2006df,King:2018fcg} there is a unique complex scalar singlet 
$\Phi$ responsible for all quark and lepton Yukawa couplings.
Moreover, in many flavon models, the top quark Yukawa coupling appears as a direct lowest order tree-level Higgs coupling as in the SM, independently of flavons, while in our model $\langle \Phi \rangle$ is indispensable for generating all Yukawa couplings, including that of the top quark. It is for such reasons that we suggest to call the associated physical Higgs singlet boson the ``Yukon'', in order to distinguish it from ``flavons'' whose phenomenology is typically very different.

The layout of the remainder of the paper is as follows. In section~\ref{model} we discuss the simplified model that we consider consisting of 
third family quarks, fourth family VLQ doublets and a gauge or global $U(1)'$ symmetry broken by a complex Higgs singlet field
$\Phi$. We show how Yukawa couplings can emerge via a seesaw-type mechanism involving mixing with the VLQs.
We also give an initial discussion of various couplings in the flavour basis, as well as the parameter space of the model.
In section~\ref{couplings} we present the couplings in the mass basis.
In section~\ref{constraints} we survey the constraints on the model.
In section~\ref{branching} we calculate the branching ratios of $T,B$ and $\phi$.
In section~\ref{collider} we consider the hadron collider signatures of $\phi$, including its production and decay cross-sections.
Section~\ref{conclusions} concludes the main body of the paper.
In Appendix~\ref{higgs} we give the full Higgs potential for the global $U(1)'$ model, as well as the Higgs mass matrices.
In Appendix~\ref{appendix:h0} we give the SM Higgs production cross-section in our model.
In Appendix~\ref{sec:appendix_decay_widths} we give the rather complicated formulae for various decay widths.

\section{The Simplified Model}
\label{model}
Although the origin of Yukawa couplings remains mysterious in the SM, certain new physics scenarios responsible for these couplings can be explored at  present and/or future hadronic machines. For example, the largest third family $(t,b)$ quark Yukawa couplings to Higgs doublets may be forbidden by a $U(1)'$ gauge or global symmetry and generated effectively after $U(1)'$ breaking via mixing with a vector-like fourth family quark doublet $Q=(T,B)$, the latter being a doublet of the SM EW group $SU(2)_W\times U(1)_Y$.  

A complete theory of this kind, capable of generating Yukawa couplings for all quarks and leptons,  
would require at least one full vector-like fourth family of fermions, $Q,U,D,L,E$, 
and an additional $U(1)'$ gauge symmetry under which the three chiral families of quarks and leptons carry zero charges \cite{King:2018fcg}. 
However, in examples of this kind \cite{King:2018fcg}, it is usually assumed that $M_4^Q\ll M_4^{D}\ll M_4^{U}$ in order to provide 
a natural explanation of small quark mixing angles.
Thus from a phenomenological point of view the lightest states $Q$ with mass $M_4^Q$ will be discovered first.
It is then possible to 
drop the heavier fourth family $SU(2)_W$ singlet quarks $U,D$ along with the vector-like leptons $L,E$, leaving us only with the single VLQ doublet $Q=(T,B)$, where vector-like means that both left (L) and right (R) chiralities have identical quantum numbers
under the SM gauge group $SU(3)_C\times SU(2)_W\times U(1)_Y$.
We also drop the first and second families of chiral SM fermions, assuming they have  interactions with neither the vector-like fermions nor the additional gauge/global symmetry. 

\begin{figure}[htbp]
	\centering
			\includegraphics[scale=0.5]{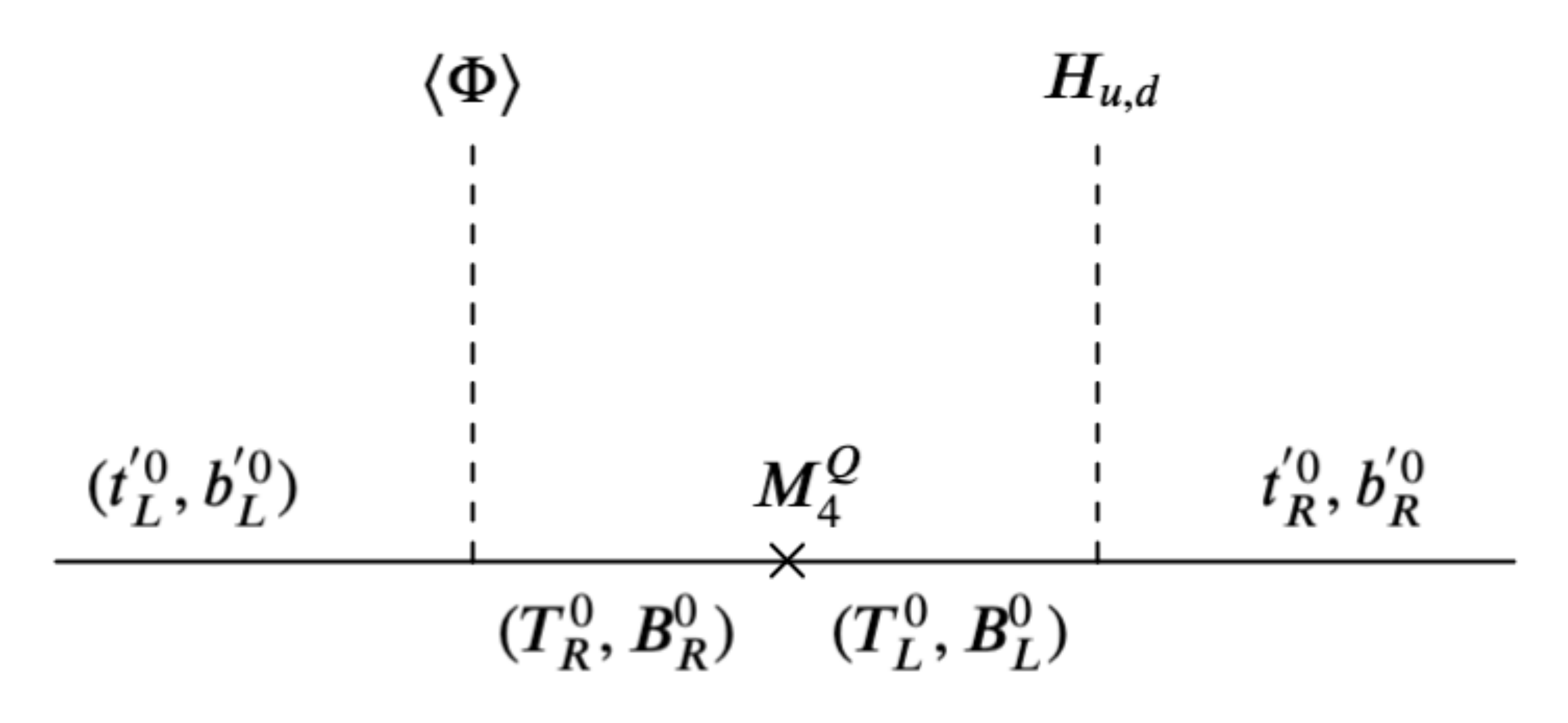}
		\caption{The origin of the third family $(t,b)$ quark Yukawa couplings in this model.}
	\label{seesaw}
\end{figure}

The resulting simplified model given in Table \ref{tab:simplified_model_reps2} includes
$Q_L$ and $Q_R$, the new vector-like doublets, together with a gauged/global $U(1)'$ which is broken by the complex Higgs singlet scalar 
$\Phi$. It also includes two Higgs doublets $H_u$, $H_d$ which are charged under $U(1)'$, preventing Yukawa couplings to the chiral quarks 
$q'_L$, $t'^0_R$, $b'^0_R$, which have no conventional Yukawa couplings.
The SM Yukawa couplings to Higgs doublets will arise only after mixing with the VLQs, as shown 
diagrammatically in a mass insertion approximation in Figure \ref{seesaw}, which is reminiscent of the seesaw mechanism.
However, due to the large top quark mass, the mass insertion approximation is not sufficient, and we need to 
use the full large angle mixing formalism introduced in \cite{King:2018fcg}, as we now discuss.

\begin{table}[htbp]
	\centering
	\renewcommand{\arraystretch}{1.2}
	\begin{tabular}{|c|c|c|c|c|c|c|c|c|}
		\hline
		Field & $q'_L$ &$t'^0_R$ & $b'^0_R$ & $Q'_L$ & $Q'_R$ & $H_u$ & $H_d$ & $\Phi$\\
		\hline
		$SU(3)_C$ & \textbf{3} &  \textbf{3} &  \textbf{3} & \textbf{3} & \textbf{3} & \textbf{1} & \textbf{1} & \textbf{1}\\
		$SU(2)_W$ & \textbf{2} &   \textbf{1} &  \textbf{1} & \textbf{2} & \textbf{2} & \textbf{2} & \textbf{2} & \textbf{1}\\
		$U(1)_Y$ & $\frac{1}{6}$ & $\frac{2}{3}$ & $-\frac{1}{3}$ & $\frac{1}{6}$ & $\frac{1}{6}$ & $\frac{1}{2}$ & $-\frac{1}{2}$ & $0$\\
		$U(1)'$ & $0$ & $0$ & $0$ & $1$ & $1$ & $-1$ & $-1$ & $1$\\
		\hline
	\end{tabular}
	\caption{Simplified vector-like fermion and $U(1)'$ gauge or global model.}
	\label{tab:simplified_model_reps2}
\end{table}
The chiral and VLQs in Table \ref{tab:simplified_model_reps2} may be written out explicitly as doublets and singlets under the SM weak gauge group $SU(2)_W$ as, 
\begin{gather}
q'_L = \begin{pmatrix}
t'^0_{L}\\
b'^0_{L}
\end{pmatrix},\quad 
t'^0_R,\quad b'^0_R,\quad 
Q'_L = \begin{pmatrix}
T^0_{L}\\
B^0_{L}
\end{pmatrix},\quad 
Q'_R =  \begin{pmatrix}
T^0_{R}\\
B^0_{R}
\end{pmatrix}.\label{eqn:definiton_of_VL_doublets}
\end{gather}
We use primes here to indicate the basis in which the fourth family $Q'_L=(T^0_L, B^0_L)$ carries unit gauged/global $U(1)'$ charges as in Table \ref{tab:simplified_model_reps2} and the third family $q'_L=(t'^0_L, b'^0_L)$ carry zero $U(1)'$  charges. In this basis the largest mass terms (ignoring contributions from SM Higgs masses) are the following:
\begin{equation}
\mathcal{L} ^{Q'} _\textrm{mass} = x_3^Q\Phi \overline{Q'_R}q'_L+M_4^Q\overline{Q}'_RQ'_L
\rightarrow \overline{Q'_R}(x_3^Q\langle \Phi \rangle {q}'_L+M_4^Q{Q}'_L)\equiv \tilde{M}_4^Q \overline{Q'_R}Q_L.
\label{eqn:heavy_mass_terms_for_VLfermions}
\end{equation}
Here, the coupling to the $\Phi$ boson that breaks $U(1)'$ is promoted to a mass term once $\Phi$ acquires a VEV, while the vector-like mass term $M_4^Q$ is a parameter of the theory.
We can combine the two terms in the bracket in Equation \eqref{eqn:heavy_mass_terms_for_VLfermions} 
into one term with a mass $ \tilde{M}_4^Q$ where we have defined a ``heavy'' field $Q_L$ (without primes) 
and an orthogonal light field $q_L$ as 
\begin{equation}
Q_L=s^Q_{34}q'_L+c^Q_{34}Q'_L,\ \ q_L=c^Q_{34}q'_L-s^Q_{34}Q'_L,
\end{equation}
where
 \begin{equation}
c^Q_{34}=\cos \theta^Q_{34},\ \   s^Q_{34}=\sin \theta^Q_{34},\ \ \tan \theta^Q_{34}=x_3^Q\langle \Phi \rangle / M_4^Q,\ \ 
(\tilde{M}_4^Q)^2=(x_3^Q)^2{\langle \Phi \rangle}^2 +  (M_4^Q)^2.
\label{4}
\end{equation}
The right-handed fields do not get rotated so we drop primes and write $Q_R=Q'_R$, $t^0_R=t'^0_R$, $b^0_R=b'^0_R$. 
In matrix notation, the results are neatly summarised as follows:
\begin{gather}
\begin{pmatrix}
q_{L} \\
Q_{L}
\end{pmatrix}
=	
\begin{pmatrix}
c^Q_{34} &  -s^Q_{34}\\
s^Q_{34} & c^Q_{34}
\end{pmatrix}
\begin{pmatrix}
q'_{L} \\
Q'_{L}
\end{pmatrix}, \ \ \ \ Q_R=Q'_R, \ \ \ \ t^0_R=t'^0_R, \ \ \ \ b^0_R=b'^0_R,
\label{eqn:decoupling_basis}
\end{gather}
where the ``light'' fields $q_L$ and the ``heavy'' fields $Q_L$ (without primes) are
\begin{gather}
q_L = \begin{pmatrix}
t^0_{L}\\
b^0_{L}
\end{pmatrix},\quad 
Q_L =  \begin{pmatrix}
T^0_{L}\\
B^0_{L}
\end{pmatrix}\,,
\label{0}
\end{gather}
and the superscript `$0$'s  denotes the states after the $\theta^Q_{34}$ rotations have been performed. Hence, these are weak eigenstates but not eigenstates of $U(1)'$, due to mixing. 

Although there are no allowed Yukawa couplings of the chiral quarks 
$q'_L$, $t'^0_R$, $b'^0_R$ to Higgs doublets, due to the gauged/global $U(1)'$, there are allowed 
Yukawa couplings which couple $t'^0_R$, $b'^0_R$ to the VLQs $Q'_L$,
\begin{align}
{\cal L}=-y^u_{43}\overline{t'^0_R}H_u Q'_L-y^d_{43}\overline{b'^0_R}H_dQ'_L+{\rm H.c.}\,.
\label{eqn:Yukawa_couplings_to_SM_Higgs}
\end{align} 
In terms of the ``light'' and ``heavy'' basis defined by Equation \eqref{eqn:decoupling_basis}, by inverting the matrix, we obtain
$Q'_L=c^Q_{34}Q_L-s^Q_{34}q_L$, which, once substituted into Equation \eqref{eqn:Yukawa_couplings_to_SM_Higgs}, will yield
effective Higgs Yukawa couplings involving ``light'' and ``heavy'' fields,
\begin{align}
{\cal L}=-y^u_{43}\overline{t'^0_R}H_u (c^Q_{34}Q_L-s^Q_{34}q_L)-y^d_{43}\overline{b'^0_R}H_d
(c^Q_{34}Q_L-s^Q_{34}q_L)
+{\rm H.c.}\,.
\label{eqn:Yukawa_couplings_to_SM_Higgs2}
\end{align} 

The full matrix of mass terms, including the effective Higgs Yukawa couplings, involving the ``light'' and ``heavy'' fields
in Equation \eqref{0} when inserted into  Equation \eqref{eqn:Yukawa_couplings_to_SM_Higgs2},
is:
\begin{gather}
{\cal L}\ \ = \ \ -\begin{pmatrix}
\overline{t}^0_{L} & 
\overline{T}^0_{L}
\end{pmatrix}
\begin{pmatrix}
-s^Q_{34}y^u_{43}\frac{v_u}{\sqrt{2}} & 0 \\
c^Q_{34}y^u_{43}\frac{v_u}{\sqrt{2}} & \tilde{M}^Q_4
\end{pmatrix}
\begin{pmatrix}
t^0_{R} \\
T^0_{R}
\end{pmatrix}
\ \ 
-
\ \ 
\begin{pmatrix}
\overline{b}^0_{L} & 
\overline{B}^0_{L}
\end{pmatrix}
\begin{pmatrix}
-s^Q_{34}y^d_{43}\frac{v_d}{\sqrt{2}} & 0 \\
c^Q_{34}y^d_{43}\frac{v_d}{\sqrt{2}} & \tilde{M}^Q_4
\end{pmatrix}
\begin{pmatrix}
b^0_{R} \\
B^0_{R}
\end{pmatrix} + \text{H.c.}\,.
\end{gather}
The matrices above show that the ``light'' and ``heavy'' states indicated by superscript `$0$'s
are not yet true mass eigenstates, since there is still some small mixing 
between them. 

The mass matrices above may be diagonalised in the small angle approximation, 
\begin{gather}
{\cal L}\ \ = \ \ -\begin{pmatrix}
\overline{t}_{L} & 
\overline{T}_{L}
\end{pmatrix}
\begin{pmatrix}
m_t & 0 \\
0 & M_T
\end{pmatrix}
\begin{pmatrix}
t_{R} \\
T_{R}
\end{pmatrix}
\ \ 
-
\ \ 
\begin{pmatrix}
\overline{b}_{L} & 
\overline{B}_{L}
\end{pmatrix}
\begin{pmatrix}
m_b & 0 \\
0 & M_B
\end{pmatrix}
\begin{pmatrix}
b_{R} \\
B_{R}
\end{pmatrix} + \text{H.c.}\,,
\label{eqn:massmatrices2}
\end{gather}
where mass eigenstates, eigenvalues and further small mixing angles are given by:
\begin{align}
	\begin{split}
		\begin{pmatrix}
		t_{L,R}\\
		T_{L,R}
		\end{pmatrix} &=\begin{pmatrix}
		\cos\theta^u_{L,R} & -\sin\theta_{L,R}^ue^{i\varphi_u} \\
		\sin\theta_{L,R}^ue^{-i\varphi_u} & \cos\theta^u_{L,R}
		\end{pmatrix}
		\begin{pmatrix}
		t^0_{L,R}\\
		T^0_{L,R}
		\end{pmatrix}, \\[2ex]
		\begin{pmatrix}
		b_{L,R}\\
		B_{L,R}
		\end{pmatrix} &=\begin{pmatrix}
		\cos\theta^d_{L,R} & -\sin\theta_{L,R}^de^{i\varphi_d} \\
		\sin\theta_{L,R}^de^{-i\varphi_d} & \cos\theta^d_{L,R}
		\end{pmatrix}
		\begin{pmatrix}
		b^0_{L,R}\\
		B^0_{L,R}
		\end{pmatrix},
	\end{split}	
\label{eqn:simple_model_mixings2}
\end{align}
\begin{equation}
m_t\approx s^Q_{34}y^u_{43}\frac{v_u}{\sqrt{2}},  \ \ M_T ^2 \approx (\tilde{M}^Q_4) ^2 + \left( \frac{c_{34} ^Q}{s_{34} ^Q} m_t \right)^2, \ \ 
\theta^u_L\approx \frac{m_t}{M_T}\theta^u_R,\ \ 
\theta^u_R\approx c^Q_{34}y^u_{43}\frac{v_u}{\sqrt{2}\tilde{M}^Q_4},
\label{eqn:approximations_for_VLT}
\end{equation}
\begin{equation}
m_b\approx s^Q_{34}y^d_{43}\frac{v_d}{\sqrt{2}},  \ \ M_B ^2 \approx (\tilde{M}^Q_4) ^2 + \left( \frac{c_{34} ^Q}{s_{34} ^Q} m_b \right)^2, \ \ 
\theta^d_L\approx \frac{m_b}{M_B}\theta^d_R,\ \ 
\theta^d_R\approx c^Q_{34}y^d_{43}\frac{v_d}{\sqrt{2}\tilde{M}^Q_4}.
\label{eqn:approximations_for_VLB}
\end{equation}
These relationships lead to the left-handed mixing angles being suppressed relative to the (small) right-handed mixing angles by ratios of third generation (SM fermion) masses to fourth family (vector-like) masses. 
The phases $\varphi_u$ and $\varphi_d$ originate from the Yukawa couplings $y^{u,d}_{43}$.
Having assumed without loss of generality that the Yukawa coupling $x_3^Q$ and mass $M_4^Q$ are real,
which can be achieved by rephasing the fermion EW doublet fields, the Yukawa couplings $y^{u,d}_{43}$ can also be made real by 
rephasing the EW singlet fields. The requirement that the mass eigenvalues be real and positive, can therefore be achieved by 
choosing the phases $\varphi_u$ and $\varphi_d$ to be $0,\pi$, which is the only remaining freedom, which however will be important later. 

\subsection{Couplings in the Flavour Basis}
The resulting couplings of the third and fourth family quarks to $W,Z$ and Higgs bosons are exactly as
given in \cite{Aguilar-Saavedra:2017giu}.
However, there will be additional couplings to $\Phi$ and (in the gauge option) $Z'$, as follows.
In the original basis the $Z'$ couples only to $Q'_L$ and $Q'_R$ (not $q'_L$ which have zero $U(1)'$ charge), like
$-g'Z'_{\mu}[\overline{Q}'_L\gamma^{\mu}Q'_L+\overline{Q}'_R\gamma^{\mu}Q'_R]$. In the decoupling basis in Equation \eqref{eqn:decoupling_basis} we have $Q'_L=c^Q_{34}Q_L-s^Q_{34}q_L$ and $Q'_R=Q_R$, so the $Z'$ couplings may be expanded as the following Lagrangians:
\begin{equation}
{\cal L}^{Z'}=  -g'Z'_{\mu}\sum_{i,j}\overline{\psi}^0_i \gamma^{\mu}({X'}_{0ij}^LP_L+{X'}_{0ij}^RP_R)\psi_j^0
\label{eqn:Zp_couplings}
\end{equation}
where the sum extends over $i,j=t,T,b,B$ and $P_{L,R}=(1\mp \gamma_5)/2$.
In the chosen basis, $t^0_L$ and $b^0_L$ carry equal $U(1)'$ charges and $t^0_R$ and $b^0_R$ have zero $U(1)'$ charges since they do not mix with $Q_R$. Couplings denoted ${X_0 ^\prime}^{L,R}_{ij}$ are given in terms of mixing angles as follows:
\begin{equation}
{X_0 ^\prime}^L_{tt}=(s^Q_{34})^2,\ \ {X_0 ^\prime}^R_{tt}=0,\ \ {X_0 ^\prime}^L_{tT}= -s^Q_{34}c^Q_{34},\ \ {X_0 ^\prime}^R_{tT}= 0,\ \ {X_0 ^\prime}^L_{TT}= (c^Q_{34})^2, \ \ {X_0 ^\prime}^R_{TT}= 1,
\label{eqn:ZptT_couplings_expanded}
\end{equation}
\begin{equation}
{X_0 ^\prime}^L_{bb}= (s^Q_{34})^2,\ \ {X_0 ^\prime}^R_{bb}= 0,\ \ {X_0 ^\prime}^L_{bB}= -s^Q_{34}c^Q_{34},\ \ {X_0 ^\prime}^R_{bB}= 0,\ \ {X_0 ^\prime}^L_{BB}= (c^Q_{34})^2, \ \ {X_0 ^\prime}^R_{BB}= 1.
\label{eqn:ZpbB_couplings_expanded}
\end{equation}
Transformation to the mass eigenstate basis requires additional small angle rotations as discussed in Equation \eqref{eqn:simple_model_mixings2}, which are straightforward to include numerically. We emphasise that in the case of the
global $U(1)'$ model there is of course no $Z'$ so all of the above couplings are absent.

The 2HDM plus Higgs singlet potential and mass matrices are given in Appendix~\ref{higgs}.
As discussed there, we shall assume that there is negligible mixing of the $\Phi$ with the two Higgs doublets, so that the physical $\phi$ Yukon predominantly arises as the real component of the complex singlet field $\Phi$ after it develops its VEV.
	This is a natural assumption in the case that $\langle \Phi \rangle $ greatly exceeds the Higgs doublet VEVs, and can be enforced by assuming certain coupling terms in the Higgs potential which couple $\Phi$ to the Higgs doublets to be small.
The $\Phi$ couplings to fermions are then obtained from rotating couplings in the $U(1)'$ flavour basis $x_3^Q\Phi \overline{q}'_LQ'_R$, after we substitute $Q'_R=Q_R$ (unrotated) and $q'_L=c^Q_{34} q_L + s^Q_{34}Q_L$ (from Equation \eqref{eqn:decoupling_basis}), yielding:
\begin{equation}
{\cal L}^{\Phi} =  - x^Q_3\Phi [	
(c^Q_{34}\overline{t}^0_L + s^Q_{34}\overline{T}^0_L)T^0_R 
+ (c^Q_{34}\overline{b}^0_L + s^Q_{34}\overline{B}^0_L)B^0_R 
]+{\rm H.c.}
\label{eqn:phi_couplings}
\end{equation}
Further rotations are required to recover couplings of the complex scalar $\Phi$, to propagating fermions. The plethora of couplings above lead to decays such as: $T\rightarrow t Z'$, $B\rightarrow b Z'$, $T\rightarrow t \phi$, $B\rightarrow b \phi$ with $Z' \rightarrow t \overline{t},b \overline{b}$ and $\phi \rightarrow t \overline{t},b \overline{b}$, where $\phi$ is the physical scalar boson 
which we assume to dominantly arise from the complex singlet field $\Phi$.

\subsection{Parameters}
In the case of the gauged model,
both Higgs doublets $H_u$ and $H_d$ couple to the $U(1)'$ sector, each with a charge of $-1$, leading to interesting phenomenology involving Higgs coupling to $Z'$. The couplings are very similar to the Higgs coupling to the usual $Z$. The SM Higgs also couples to the $\phi$ boson. 

We now derive the parameters of interest for later calculation of relevant observables including mass eigenvalues of the SM top and vector-like top (VLT) in terms of model parameters, making use of Equations \eqref{4}, \eqref{eqn:approximations_for_VLT} and \eqref{eqn:approximations_for_VLB}, as well as $M_{Z'} = g' v_\phi$.
We may rewrite all of these quantities without loss of generality in terms of the parameter set 
for the gauge $U(1)'$ model of: $\{ M_T ,~ M_{Z'} ,~ M_\phi ,~ g' ,~ \theta ^Q _{34} ,~ \tan \beta \}$ and SM quantities. This is one possible choice of many equivalent descriptions. Note that for $(x_3 ^Q ,~ M_4 ^Q)$ we have solved the system of simultaneous equations and used the trigonometric identity $\sin(x) = \sqrt{\tan ^2 (x) / 1+ \tan ^2 (x)}$ so that;
\begin{equation}
y^u_{43}\approx \frac{\sqrt{2} m_t}{s^Q_{34} v \sin \beta},  \ \ (\tilde{M}^Q_4 )^2 \approx M_T ^2 - \left( \frac{c_{34} ^Q}{s_{34} ^Q} m_t \right)^2 ,
\label{eqn:derived_params_1}
\end{equation}
\begin{equation}
y^d_{43}\approx \frac{\sqrt{2} m_b}{s^Q_{34} v \cos \beta},  \ \ (M_B)^2 \approx M_T ^2 - \left( \frac{c_{34} ^Q}{s_{34} ^Q} \right) ^2 ( m_t ^2 - m_b ^2 ) , \ \ 
\label{eqn:derived_params_2}
\end{equation}
\begin{equation}
\theta^u_L\approx \frac{m_t}{M_T}\theta^u_R,\ \ 
\theta^u_R\approx c^Q_{34}y^u_{43}\frac{v \sin \beta}{\sqrt{2} \, M_T} = \frac{c_{34}^Q m_t}{s_{34} ^Q M_T},
\label{eqn:derived_params_1_2}
\end{equation}
\begin{equation}
\theta^d_L\approx \frac{m_b}{M_B}\theta^d_R, \ \ 
\theta^d_R\approx c^Q_{34}y^d_{43}\frac{v \cos \beta}{\sqrt{2} \, M_B} = \frac{c_{34}^Q m_b}{s_{34} ^Q M_T}, \ \ 
\label{eqn:derived_params_3}
\end{equation}
\begin{equation}
v_\phi = \frac{M_{Z'}}{g'} ,\ \  {x_3 ^Q} = \sin(\theta _{34} ^Q) \frac{M_T}{\langle \Phi \rangle} , \ \ {M_4 ^Q} = M_T \cos \theta _{34} ^Q .
\label{eqn:derived_params_4}
\end{equation}

In the case of the global model similar results apply but with the smaller parameter set:
$\{ M_T ,~ v_{\phi} ,~ M_\phi ,~ \theta ^Q _{34} ,~ \tan \beta \}$ and SM quantities. 
Note that in the case $g'=1$ (as we assume in this paper) $v_{\phi}=M_{Z'}$. 
Thus many of the results that we present later for a particular value $Z'$ of mass in the gauge model are 
equally valid for the case of the global model with a $v_{\phi}$ of the same value.
Also note that  $\langle \Phi \rangle= v_{\phi}/\sqrt{2}$.

\section{Couplings in the Mass Basis}
\label{couplings}
We now present the full couplings in the mass basis. The content involving the $Z'$ and $\phi$ is new whereas the remaining couplings to $W,~Z$ and $h^0$ (the aforementioned SM-like Higgs state) are taken from \cite{Aguilar-Saavedra:2017giu}. Here, $h^0$ is the lightest Higgs state with SM-like couplings and the other Higgs states are heavier and decoupled.

\subsection{Interactions Between Light and Heavy States}
The following interactions will determine the decays of the new heavy VL quark states, as well as for the Yukon $\phi$. The interactions with the SM gauge states $W$ and $Z$ are the same as in the usual one Higgs doublet model, and 
\begin{align}
	\mathcal{L}_W & =  -\frac{g}{\sqrt 2} \left[
	\bar{T}_L \gamma ^\mu V^L_{Tb} b_L + \bar{T}_R \gamma ^\mu V^R _{Tb} b_R +
	\bar{t}_L \gamma ^\mu V_{tB} ^L B_L + \bar{t}_R \gamma ^\mu V^R _{tB} B_R
	\right] \Wm^+ +\text{H.c.} \,, \notag \\[1mm]
	\mathcal{L}_Z & =  -\frac{g}{2 c_W} \left[ \bar{t}_L \gamma ^\mu X_{tT}^LT_L +  \bar{t}_R \gamma ^\mu X_{tT}^R T_R
	- \bar{b}_L \gamma ^\mu X_{bB}^L B_L - \bar{b}_R \gamma ^\mu X_{bB}^R B_R \right] \Zm +\text{H.c.} \,, \notag \\[1mm]
	\mathcal{L}_{Z'} & =  -g' \left[ \bar{t}_L \gamma ^\mu {X'}_{tT}^L T_L +  \bar{t}_R \gamma ^\mu {X'}_{tT}^R T_R
	- \bar{b}_L \gamma ^\mu {X'}_{bB}^L B_L - \bar{b}_R \gamma ^\mu {X'}_{bB}^R B_R \right] {Z'}_\mu +\text{H.c.}\,.
\end{align}
Where the couplings as a function of the quark mixing angles and phases are 
\begin{align}
	&   V_{Tb}^L = {\sin \theta}_L^u \cos \theta_L^d e^{-i \varphi_u} - {\cos \theta}_L^u {\sin \theta}_L^d e^{- i \varphi_d} \,,
	&& V_{Tb}^R = -{\cos \theta}_R^u {\sin \theta}_R^d e^{- i \varphi_d} \,, \notag \\
	&   V_{tB}^L = {\cos \theta}_L^u {\sin \theta}_L^d e^{i \varphi_d} - {\sin \theta}_L^u {\cos \theta}_L^d e^{i \varphi_u} \,, 
	&& V_{tB}^R = -{\sin \theta}_R^u {\cos \theta}_R^d e^{i \varphi_u} \,, \notag \\
	&   X_{tT}^L = 0 \,,
	&& X_{tT}^R = -{\sin \theta}_R^u {\cos \theta}_R^u e^{i \varphi_u} \,, \notag \\
	&   X_{bB}^L = 0 \,,
	&& X_{bB}^R = -{\sin \theta}_R^d {\cos \theta}_R^d e^{i \varphi_d} \,, \notag \\
	&   {X'}_{tT}^L = 
(s_{34}^Q {\cos \theta}_L^u+e^{i \varphi _u} c_{34}^Q {\sin \theta}_L^u) (-c_{34}^Q {\cos \theta}_L^u+e^{i \varphi _u} s_{34}^Q {\sin \theta}_L^u)
	\,,
	&& {X'}_{tT}^R = -{\sin \theta}_R^u {\cos \theta}_R^u  e^{i \varphi _u} \,, \notag \\
	&   {X'}_{bB}^L =(s_{34}^Q {\cos \theta}_L^d+e^{i \varphi _d} c_{34}^Q {\sin \theta}_L^d) (-c_{34}^Q {\cos \theta}_L^d+e^{i \varphi _d} s_{34}^Q {\sin \theta}_L^d) \,,
	&& {X'}_{bB}^R =- {\sin \theta}_R^d {\cos \theta}_R^d  e^{i \varphi _d} \,.
\end{align}
Here and throughout, we work in the so-called alignment limit for the 2HDM, $\beta - \alpha = \pi/2$, such that interactions with the lightest scalar $h^0$ are the same as with one Higgs doublet, as shown below together with the Yukon $\phi$ couplings,
\begin{align}
	\mathcal{L}_{h^0} & = -\frac{g M_T}{2M_W} \left( \bar{t}_R Y_{tT}^L T_L +  \bar{t}_L Y_{tT}^R T_R \right)  h^0 
	-\frac{g M_B}{2M_W} \left( \bar{b}_R Y_{bB}^L B_L +  \bar{t}_L Y_{bB}^R B_R \right) h^0 +\text{H.c.} \,, \notag \\
	\mathcal{L}_{\phi} & = -\frac{g' M_T}{M_{Z'}}  \left( \bar{t}_R {Y'}_{tT}^L T_L +  \bar{t} _L {Y'}_{tT}^R T_R \right) \phi 
	-\frac{g' M_T}{M_{Z'}} \left( \bar{b}_R {Y'}_{bB}^L B_L +  \bar{b}_L {Y'}_{bB}^R B_R \right)  \phi +\text{H.c.},
\end{align}
where the couplings are
\begin{align}
	&   Y_{tT}^L = {\sin \theta}_R^u {\cos \theta}_R^u e^{i \varphi_u} \,,
	&& Y_{tT}^R = \frac{m_t}{M_T} {\sin \theta}_R^u {\cos \theta}_R^u e^{i \varphi_u} \,, \notag \\
	&   Y_{bB}^L = {\sin \theta}_R^d {\cos \theta}_R^d e^{i \varphi_d} \,,
	&& Y_{bB}^R = \frac{m_b}{M_B} {\sin \theta}_R^d {\cos \theta}_R^d e^{i \varphi_d} \,, \notag \\
	&   {Y'}_{tT}^L = -e^{i \varphi _u} s_{34}^Q {\sin \theta}_R^u (s_{34}^Q {\cos \theta}_L^u+e^{i \varphi _u} c_{34}^Q {\sin \theta}_L^u) \,,
	&& {Y'}_{tT}^R = s_{34}^Q {\cos \theta}_R^u (c_{34}^Q {\cos \theta}_L^u-e^{i \varphi _u} s_{34}^Q {\sin \theta}_L^u) \,, \notag \\
	&   {Y'}_{bB}^L =-e^{i \varphi _d} s_{34}^Q {\sin \theta}_R^d (s_{34}^Q {\cos \theta}_L^d+e^{i \varphi _d} c_{34}^Q {\sin \theta}_L^d) \,,
	&& {Y'}_{bB}^R = s_{34}^Q {\cos \theta}_R^d (c_{34}^Q {\cos \theta}_L^d-e^{i \varphi _d} s_{34}^Q {\sin \theta}_L^d) \,.
\end{align}


\subsection{Interactions Between Light and Light States}
The usual light SM-like quark states will have modified gauge and scalar interactions to the SM content, in addition to new interactions with the $U(1)'$ gauge boson. Constraints on the SM interactions are explored in \cite{Dawson:2012di,Aguilar-Saavedra:2013qpa} and we will not discuss them here. The new interactions are as follows,
\begin{align}
	\mathcal{L}_W & = -\frac{g}{\sqrt 2} \left[ \bar{t}_L \gamma ^\mu  V_{tb}^L b_L + t_R  \gamma ^\mu V_{tb}^R b_R \right] \Wm^+ +\text{H.c.} \,, \notag \\[1mm]
	\mathcal{L}_Z & = -\frac{g}{2 c_W} \left[ \bar{t}_L \gamma ^\mu X_{tt}^L t_L + \bar{t}_R \gamma ^\mu X_{tt}^R t_R - \bar{t} \gamma ^\mu (2  Q_t s_W^2) t - \bar{b}_L \gamma ^\mu  X_{bb}^L b_L - \bar{b}_R \gamma ^\mu X_{bb}^R t_R -  \bar{b} \gamma ^\mu (2  Q_b s_W^2) b \right] Z_\mu \,,  \notag \\[1mm]
	\mathcal{L}_{Z'} & = - g' \left[ \bar{t}_L \gamma ^\mu {X'}_{tt}^L t_L + \bar{t}_R \gamma^\mu {X'}_{tt}^R t_R - \bar{b}_L \gamma ^\mu {X'}_{bb}^L b_L - \bar{b}_R \gamma ^\mu {X'}_{bb}^R b_R  \right] Z' _\mu \,,
\end{align}
with the couplings as
\begin{align}
	&   V_{tb}^L = {\cos \theta}_L^u {\cos \theta}_L^d + {\sin \theta}_L^u {\sin \theta}_L^d e^{i (\varphi_u - \varphi_d)} \,,
	&& V_{tb}^R = {\sin \theta}_R^u {\sin \theta}_R^d e^{i (\varphi_u - \varphi_d)} \,, \notag \\
	&   X_{tt}^L = 1 \,,
	&& X_{tt}^R = {\sin^2 \theta}_R^{u} \,, \notag \\
	&   X_{bb}^L = 1 \,,
	&& X_{bb}^R = {\sin^2 \theta}_R^{d} \,. \notag \\
	&   {X'}_{tt}^L =2 c_{34}^Q s_{34}^Q \cos (\varphi _u) {\cos \theta}_L^u {\sin \theta}_L^u+(s_{34}^Q)^2 ({\cos \theta}_L^u)^2+(c_{34}^Q)^2 ({\sin \theta}_L^u)^2 \,,
	&& {X'}_{tt}^R ={{\sin^2 \theta}_R^u} \,, \notag \\
	&   {X'}_{bb}^L =2 c_{34}^Q s_{34}^Q \cos (\varphi _d)  {\cos \theta}_L^d {\sin \theta}_L^d+(s_{34}^Q)^2 ({\cos \theta}_L^d)^2+(c_{34}^Q)^2 ({\sin^2 \theta}_L^d) \,,
	&& {X'}_{bb}^R = {{\sin^2 \theta}_R^d} \,.
\end{align}
Given the small size of the mixing angles, these are quite close to the usual SM interactions. The scalar interactions between light quarks and Higgs fields are as follows;
\begin{align}
	\mathcal{L}_{h^0} & = -\frac{g m_t}{2 M_W} Y_{tt} \; \bar t \, t \, h^0 
	-\frac{g m_b}{2M_W} Y_{bb} \; \bar b \, b \, h^0 \,, \notag \\
		\mathcal{L}_{\phi} & = -\frac{g' M_T}{M_{Z'}} {Y'}_{tt} \; \bar t \, t \, \phi 
	-\frac{g' M_T}{M_{Z'}} {Y'}_{bb} \; \bar b \, b \, \phi \,.
	\label{29}
\end{align}
with the couplings as
\begin{align}
	&   Y_{tt} = {\cos^2 \theta}_R^{u} \,,
	&& Y_{bb} = {\cos^2 \theta}_R^{d} \,,  \notag \\
	&   {Y'}_{tt} = s_{34}^Q {\sin \theta}_R^u (s_{34}^Q {\sin \theta}_L^u-e^{i \varphi _u} c_{34}^Q {\cos \theta}_L^u) \,,
	&& {Y'}_{bb} = s_{34}^Q {\sin \theta}_R^d (s_{34}^Q {\sin \theta}_L^d-e^{i \varphi _d} c_{34}^Q {\cos \theta}_L^d) \,.
\label{30}
\end{align}
We will explore the effect of deviations in $Y_{tt}$ in section \ref{sec:constraints_higgs}.


\subsection{Interactions Between Heavy and Heavy States}
For the heavy VLQ interactions we find the following,
\begin{align}
	\mathcal{L}_W & =  -\frac{g}{\sqrt 2} \left[
	\bar{T}_L \gamma ^\mu V^L_{TB} B_L + \bar{T}_R \gamma^\mu V^R _{TB} B_R +
	\bar{t}_L \gamma^\mu V_{TB} ^L B_L + \bar{T}_R \gamma ^\mu V^R _{TB} B_R
	\right] \Wm^+ +\text{H.c.} \,, \notag \\[1mm]
	\mathcal{L}_Z & =  -\frac{g}{2 c_W} \left[ \bar{T}_L \gamma ^\mu X_{TT}^LT_L +  \bar{T}_R \gamma ^\mu X_{TT}^R T_R
	- \bar{B}_L \gamma ^\mu X_{BB}^L B_L - \bar{B}_R \gamma ^\mu X_{BB}^R B_R \right] \Zm +\text{H.c.} \,, \notag \\[1mm]
	\mathcal{L}_{Z'} & =  -g' \left[ \bar{T}_L \gamma ^\mu {X'}_{TT}^L T_L +  \bar{T}_R \gamma ^\mu {X'}_{TT}^R T_R
	- \bar{B}_L \gamma^\mu {X'}_{BB}^L B_L - \bar{B}_R \gamma ^\mu {X'}_{BB}^R B_R \right] {Z'}_\mu +\text{H.c.} \,, \notag \\[1mm]
		\mathcal{L}_{h^0} & = -\frac{g m_t}{2M_W} Y_{TT} \; \bar T \, T \, h^0 
	-\frac{g m_b}{2M_W} Y_{BB} \; \bar B \, B \, h^0 \,, \notag \\
	\mathcal{L}_{\phi} & = -\frac{g' M_T}{M_{Z'}} {Y'}_{TT} \; \bar T \, T \, \phi 
	-\frac{g' M_T}{M_{Z'}} {Y'}_{BB} \; \bar B \, B \, \phi \, .
\end{align}
The heavy-heavy interaction Lagrangians are similar the the above cases, but with largely replacing $t\rightarrow T$ and $b\rightarrow B$ (noting the pre-factor for $h^0$ interactions remains the top mass $m_t$).
\begin{align}
	&   V_{TB}^L = {\cos \theta}_L^u {\cos \theta}_L^d + {\sin \theta}_L^u {\sin \theta}_L^d e^{-i (\varphi_u - \varphi_d)} \,,
	&& V_{TB}^R = {\cos \theta}_R^u {\cos \theta}_R^d  \,, \notag \\
	&   X_{TT}^L = 1 \,,
	&& X_{TT}^R = {\cos^2 \theta}_R^{u} \,, \notag \\
	&   X_{BB}^L = 1 \,,
	&& X_{BB}^R = {\cos^2 \theta}_R^{d} \,, \notag \\
	&   {X'}_{TT}^L = \frac{1}{2} (-4 c_{34}^Q s_{34}^Q \cos (\varphi _u) {\cos \theta}_L^u {\sin \theta}_L^u+\cos (2 \theta _{34}^Q) \cos (2 \theta _L^u)+1) \,,
	&& {X'}_{TT}^R = {\cos^2 \theta}_R^u \,, \notag \\
	&   {X'}_{BB}^L = \frac{1}{2} (-4 c_{34}^Q \cos (\varphi _d) s_{34}^Q {\cos \theta}_L^d {\sin \theta}_L^d+\cos (2 \theta _{34}^Q) \cos (2 \theta _L^d)+1) \,,
	&& {X'}_{BB}^R ={\cos^2 \theta}_R^d \,, \notag \\
	&   Y_{TT} = {\sin \theta}_R^{u\,2} \,,
	&& Y_{BB} = {\sin^2 \theta}_R^{d} \,, \notag \\
	&   {Y'}_{TT} = (s_{34}^Q)^2 {\cos \theta}_R^u ({\cos \theta}_L^u+e^{i \varphi _u} {\sin \theta}_L^u \cot (\theta _{34}^Q)) \,,
	&& \notag \\
	&   {Y'}_{BB} = (s_{34}^Q)^2 {\cos \theta}_R^d ({\cos \theta}_L^d+e^{i \varphi _d} {\sin \theta}_L^d \cot (\theta _{34}^Q)) \,.
\end{align}

\section{Constraints}
\label{constraints}
Before considering the phenomenology of the new fields in this model, we realise that we do not have complete freedom in our entire parameter set $\{ M_T ,~ M_{Z'} ,~ M_\phi ,~ g' ,~ \theta ^Q _{34} ,~ \tan \beta \}$. Thus, we test to what extent the production of the SM Higgs is altered by the presence of mixing between the SM-like and VLT. We then focus on constraints arising from Flavour Changing Neutral Currents (FCNCs) and EW Precision Observables (EWPOs) affecting the Z'. Finally, there are theoretical constraints from perturbativity to account for too.

\subsection{Standard Model Higgs Production}
\label{sec:constraints_higgs}
\begin{figure}[htbp]
	\centering
	\begin{subfigure}{0.45\textwidth}
		\includegraphics[scale=1.2]{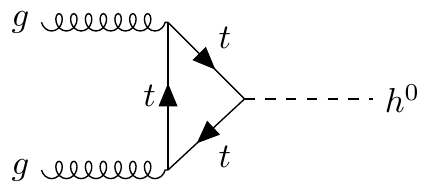}
	\end{subfigure}
	\begin{subfigure}{0.45\textwidth}
		\includegraphics[scale=1.2]{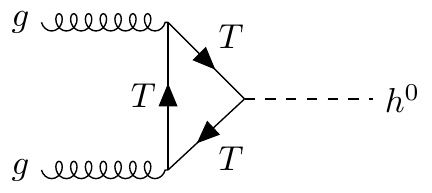}
	\end{subfigure}
	\caption{SM Higgs production dominant diagrams at the LHC.}
	\label{fig:SM_Higgs_production_gluons}
\end{figure}

The presence of additional quark content in this model can alter the production cross-section of the SM Higgs boson at the LHC. In Figure \ref{fig:SM_Higgs_production_gluons}, we draw the Feynman diagrams of dominant interactions. In principle the LHC is sensitive to deviations of $Y_{tt}= \cos^2 \theta_R^u$ by a few percent, but any deviation here is largely compensated by the contribution from $Y_{TT}=\sin^2 \theta_R^u$. We present a full derivation of the production cross-section in Appendix \ref{appendix:h0} and summarise the result here. We include the dominant contributions from third generation top and VLT, $\{t,~T\}$, yielding:
\begin{align}
	\sigma(gg\rightarrow h^0)|_{{\rm NLO}}=27.55~\text{pb} ,
\end{align}
for $\{\sqrt{s}=14~\text{TeV},~M_T=1.5~\text{TeV},~\thQ =\frac{\pi}{4}\}$. We calculated the cross-section at Next-to-Leading Order (NLO) in QCD via a $k$-factor of 1.7 \cite{Spira:1997dg}. This NLO cross-section is within the SM NLO errors \cite{Aaboud:2018ezd,GomezBock:2007hp}, so that it presents no constraints or discovery potential
\footnote{To match the observed LHC cross-sections (see e.g. \cite{Aaboud:2018ezd}) for $\sigma(gg\rightarrow h^0)$, one requires larger loop orders than NLO, such as N3LO. In this work we fix NLO in the signal for consistency to match the backgrounds for $\gamma \gamma ,~tth,~ttZ$, which are only known to NLO.}
.  The same conclusions can be obtained for the other choices of VLQ parameters that we will use in the remainder of the numerical analysis.

\subsection{$Z'$ Constraints}
Strict constraints apply to the new $Z'$ vector boson from FCNCs plus EWPOs and we take very conservative limits here.

\subsubsection{FCNC Constraints}
Since the new vector boson has family non-universal couplings, this results in quark FCNCs from the coupling~\cite{King:2018fcg}
\footnote{As discussed in detail in~\cite{King:2018fcg}, the $Z'$ induces non-universal and flavour-violating couplings due to the fact that the chiral quarks and leptons do not carry $U(1)'$ charges, while the vector-like fermions do, and hence the light mass eigenstates will couple to the $Z'$ according to the degree of mixing with vector-like fermions. There will be also $Z$ induced FCNCs due to mixing with vector-like fermions but these will be suppressed by the ratio of fermion masses to heavy vector-like fermion masses.}
:
\begin{equation}
\lagr \supset Z' _\mu (g_{bs} \bar{s}_L \gamma ^\mu b_L),
\end{equation}
where;
\begin{equation}
g_{bs} \approx g' (s^Q _{34})^2 V_{ts}.
\end{equation}
This effect has been studied previously and one finds a bound~\cite{King:2018fcg},
\begin{equation}
	M_{Z'} \gtrsim g' (s_{34} ^Q) ^2 (6 \textrm{ TeV}).
	\label{eqn:Steve_Z'_bound}
\end{equation}
To maximise Yukawa couplings, which are proportional to the inverse of $v_\phi$, we thus wish to maximise the ratio $R({g'}) \equiv g' / M_{Z'}$. Here, we find a linear scaling and so no preferred value of the gauge coupling.

\subsubsection{EWPO Constraints}
There is a model-dependent limit \cite{Coriano:2015sea,Accomando:2016sge} from EWPO data  at LEP-II, despite the requirement of the absence of direct couplings between SM fermions and the new $U(1)'$ gauge boson. The diagonalisation of the mass matrix extracted from the $U(1)'$ extended Lagrangian  gives relations
between mass ($A^\mu, Z^\mu$ and $Z^{\prime\mu}$) and interaction  ($B^\mu, W_3^\mu$ and $B^{\prime\mu}$) eigenstates in the neutral EW sector. This introduces, alongside 
the usual  Weinberg angle of the SM, also a new mixing angle $\theta^\prime$ between the $Z$ and $Z'$, where $-\pi/4 \leq\theta^\prime \leq +\pi/4$.
A bound on the mixing angle $\theta^\prime$ has been obtained in Ref.~\cite{Abreu:1994ria}, which constrains it to small values, namely, $ |\theta^\prime|\lesssim
10^{-3}$. Over such a range, $\theta'$   can be written approximately as follows:
\begin{equation}
	\theta ' \approx  g' \frac{M_Z ^2}{M_{Z'} ^2 - M_Z ^2},
\end{equation}
so that one obtains
\begin{equation}
 M_{Z'}^2 \gtrsim M_Z ^2 \left( \frac{g'+\theta'}{\theta'} \right).
\end{equation}
Unlike in the FCNC case, the ratio $R(g')$ no longer linearly scales and is instead optimised for larger values of the gauge coupling. Taking both FCNC and EWPO constraints into account, we may find the optimal gauge coupling to maximise the allowed value of $R(g')$, which is $g' \approx 0.93$ and thus $M_{Z'} \approx 2775$ GeV. However, here, and throughout, for simplicity, we will take the value of $g'=1$. This value has the additional benefit of aligning the VEV of $\Phi$ with the $Z'$ mass, $M_{Z'}=v_\phi$. In this case we find the two constraints to be
	\begin{align}
		\textrm{FCNCs:}~~~&M_{Z'} \gtrsim 3000~\textrm{GeV},  \\
		\textrm{EWPOs:}~~~&M_{Z'} \gtrsim 2885~\textrm{GeV},  
			\end{align}
and so take the stronger FCNC requirement, by fixing throughout $g'=1$ and $M_{Z'}=3000$ GeV.

\subsection{Constraints on the Higgs singlet Yukon $\phi$}
The same FCNC limits apply to the Yukon $\phi$, however, they place much weaker bounds than for the $Z'$ due to the small coupling between $\phi$ and the SM $b$ quark. 
The mass basis couplings between SM quarks and $\phi$
from Equations \eqref{29} and \eqref{30} may be written in a simpler form as 
\begin{equation}
	-{\cal L} ^\phi =  \tilde{Y}' _{bb}  \bar{b}_L \phi b_R  + \tilde{Y}' _{tt}  \bar{t}_L \phi t_R + {\rm H.c.},
	\label{YukonLag}
\end{equation}
where we have defined $ \tilde{Y}' _{bb} \equiv \frac{g' M_T}{M_{Z'}} {Y'}_{bb} $ and $\tilde{Y}' _{tt}\equiv \frac{g' M_T}{M_{Z'}} {Y'}_{tt}$.
In terms of these couplings, one finds the following limit from $B$ meson oscillations \cite{Artuso:2015swg}:
\begin{equation}
	M_{\phi} > |\tilde{Y}' _{bb}| (6 \textrm{ TeV}),
\end{equation}
where, using the small angle approximation, the leading order term of this coupling is
\begin{equation}
	|\tilde{Y}' _{ff}| \simeq {(c_{34} ^Q)}^2 \frac{m_f}{v_{\phi}}. 
	\label{Yukoncoupling}
\end{equation}
These small couplings lead to very weak constraints. For example, with $v_{\phi} = 1000$ GeV, $c_{34} ^Q = 1/\sqrt{2}$, one finds $M_\phi \gtrsim 10 $ GeV. Note that Equation \eqref{Yukoncoupling}
clearly shows that the characteristic Yukon coupling to fermions is proportional to the fermion mass, which results from its involvement in the origin of fermion Yukawa couplings.
\subsection{Theoretical Constraints}
\subsubsection{Perturbativity Constraints on $\theta ^Q _{34}$}
Requiring that the Yukawa coupling of the top quark is perturbative, we find, with $\tan \beta =30$, the following constraint:
\begin{equation}
	y^u_{43} = \frac{\sqrt{2} m_t}{s^Q_{34} v \sin \beta} \gtrsim 2 ~~\longrightarrow~~ \theta ^Q _{34} \gtrsim 0.52 \sim \pi/6.
\end{equation}
Where possible, in this paper, we will present results for both $\theta ^Q _{34}=\pi/4$ (the most natural case) and the perturbative limit scenario, $\theta ^Q _{34}=\pi/6$.
\subsubsection{Perturbativity Constraints on $\lambda_6$}
In the CP-even mass matrix listed in Appendix \ref{subsec:mass_mamtrix_for_CP_even_odd_charged}, there is a limit on the quartic potential term for $\lambda$ to be perturbative. Assuming small mixing in the potential, we find
\begin{equation}
	M_\phi ^2 \approx 2 \lambda_6 v_\phi^2 ,
	\end{equation}
so that by placing a conservative limit on $\lambda_6 \lesssim 2$ one finds
\begin{equation}
	v_\phi \gtrsim \frac{M_\phi}{2}.
	\label{eq:vev_phi}
	\end{equation}

\section{Branching Ratios of the VLQs $T,B$ and the Yukon $\phi$}
\label{branching}
\subsection{$T$ and $B$ Decays}
After decoupling the additional 2HDM content and with a too heavy (kinematically forbidden) $Z'$, there are four possible decay modes for the heavy VLQs $T$ and $B$,
\begin{align}
	T &\rightarrow h^0 t ,~~~ T \rightarrow Zt ,~~~ T \rightarrow Wb ,~~~ T \rightarrow \phi t ,\\
	B &\rightarrow h^0 b ,~~~ B \rightarrow Zb ,~~~ B \rightarrow Wt ,~~~ T \rightarrow \phi b	,
	\end{align}
where the expressions for the various decay modes are given in Appendix \ref{sec:appendix_decay_widths}. 

\begin{figure}[!t]
	\centering
	\includegraphics[width=0.75\linewidth]{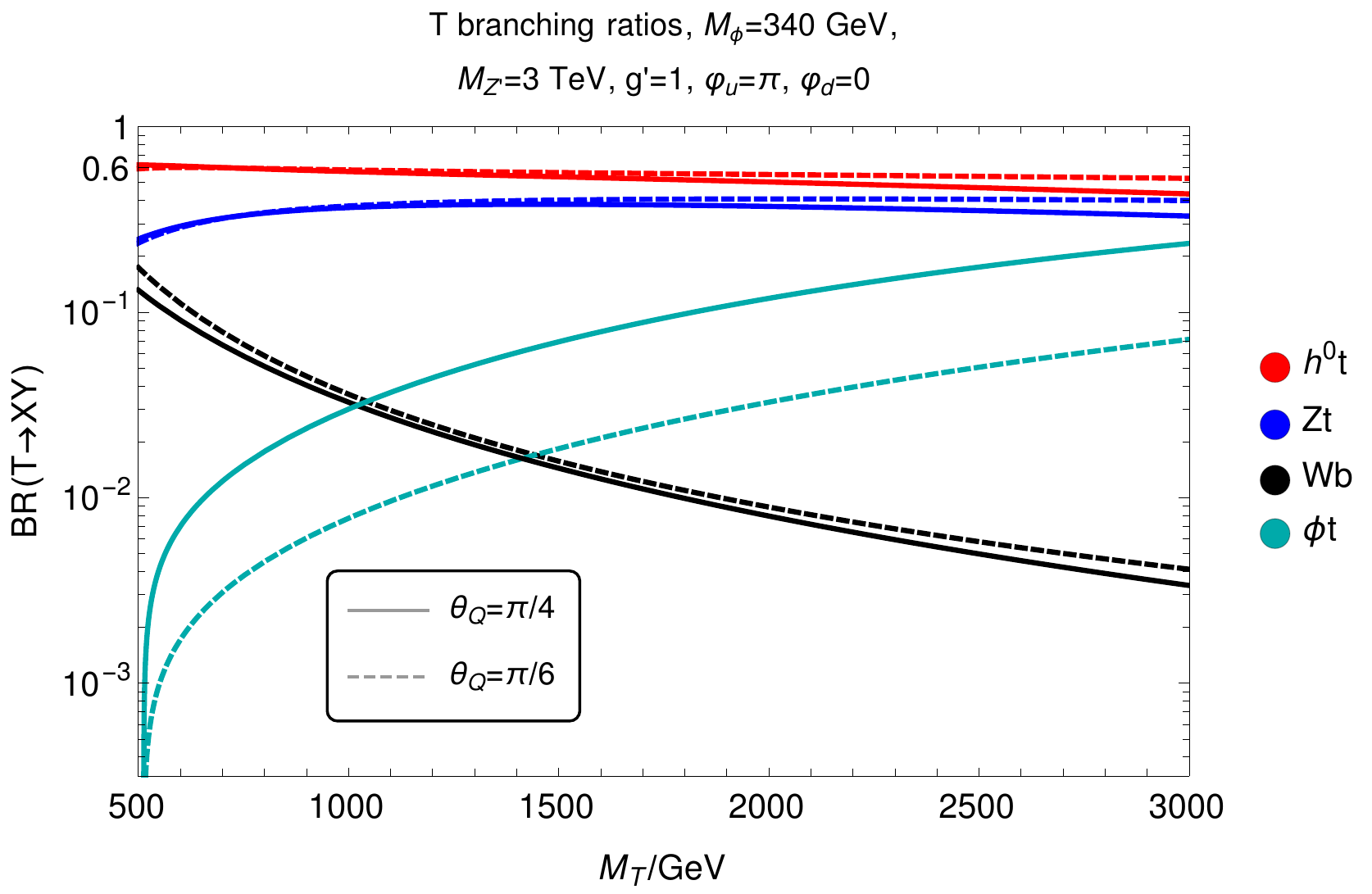}\\[0.3cm]
	\includegraphics[width=0.75\linewidth]{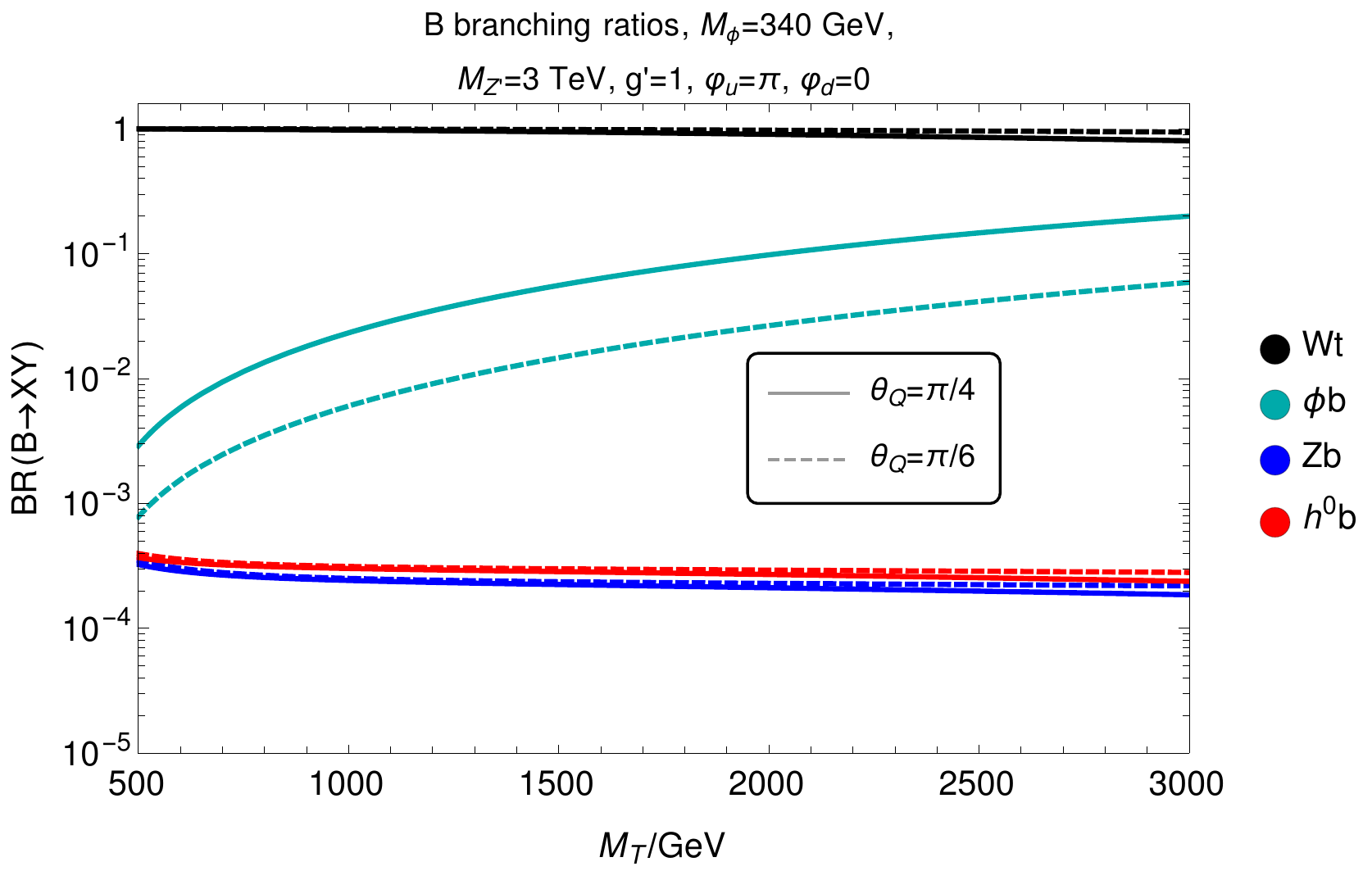}
	\caption{BRs of $T$ (upper) and $B$ (lower) as a function of the VLT mass for $\theta ^Q _{34}= \pi /4$ drawn in solid lines and $\theta ^Q _{34}= \pi /6$ drawn in dashed lines. 
}
	\label{fig:plottbrthetaq}
\end{figure}

We present the results of these Branching Ratios (BRs) of the VLT in Figure \ref{fig:plottbrthetaq}. We fix the $Z'$ mass to be $M_{Z'}=3000$ GeV and the gauge coupling to be $g'=1$, as discussed previously, while choosing the phase regime $\varphi _u = \pi ,~ \varphi _d =0$, which will be motivated later. We further choose an example $\phi$ mass to be smaller than the $T,~B$ quark ones (as an example, we have taken $M_{\phi}=340$ GeV) as an attempt to maximise its production via VLQ decay. We can see that, for the VLT, the $T \rightarrow h^0 t$ and $T \rightarrow Zt$ modes dominate and for the Vector-Like Bottom (VLB) the $B \rightarrow Wt$ mode dominates. Given the large BRs to SM content, we may straightforwardly extract the experimental limit on the VLT mass as $M_T \gtrsim 1$ TeV, from which we derive a VLB mass of $M_B = 956$ GeV, which is stronger than the independently calculated VLB experimental limit \cite{Aguilar-Saavedra:2017giu}. Henceforth, we will use these two masses as our standard parameter set. Since the BR to $\phi$ is very small in both cases, it is not possible to attempt any discovery with this mass spectrum. Instead, we will focus in the remainder of the paper on the scenario where $M_{\phi} > M_{T}$. Numerically, in this case we find, with $M_T=1000$ GeV, with negligible difference between $\theta ^Q _{34} = \pi /4$ and $\pi/6$, that
\begin{align}
	\textrm{BR}(T \rightarrow h^0 t)&=0.589, \\ 
	\textrm{BR}(T \rightarrow Z t)&=0.377, \\
	\textrm{BR}(B \rightarrow Wt)&=0.999. \\
	\end{align}

\subsection{Higgs Singlet Yukon $\phi$ Decays}
There are several possible decay modes of the Yukon $\phi$. After decoupling the $Z'$ and 2HDM content and requiring a not overly heavy $\phi$, $M_{\phi}<2 M_B$, we are left with the following two-body decay modes:
\begin{alignat}{3}
	&\phi \rightarrow bb ,~~~ &&\phi \rightarrow tt ,~~~ &&\phi \rightarrow tT ,~~~ \phi \rightarrow bB, \\
	&\phi \rightarrow \gamma \gamma , &&\phi \rightarrow gg , &&\phi \rightarrow Z \gamma,
\end{alignat}
where the full partial width expressions are derived in Appendix \ref{sec:appendix_decay_widths}. In Figure \ref{fig:phi_BRs}, we plot the BRs as a function of the scalar mass, $M_{\phi}$, for both $\thQ=\pi /4$ in solid colours and $\thQ=\pi/6$ in dashed ones. Some explanation is required for the peculiar behaviour of $\phi \rightarrow Z \gamma$. Unlike for $\gamma$, the $Z$ can couple also non-diagonally SM quarks to VLQs, so the $Z \gamma$ partial width has six amplitude contributions from loops of $tt,~TT,~tT,~bb,~BB,~bB$. As these six amplitudes smoothly change with $M_{\phi}$, the sum of these $\sum _i A_i$ smoothly changes above and below zero several times with low $\phi$ masses. Since the partial width is the absolute square of the sum of the amplitudes, as the sum of  amplitudes transitions from negative to positive, the partial width produces a cancellation to zero and then increases again, causing the oscillating behaviour. For the $\thQ=\pi/6$ case, the sum of the amplitudes does not cross the axis, and no such a behaviour is seen.

\begin{figure}[!t]
	\centering
	\includegraphics[width=0.75\textwidth]{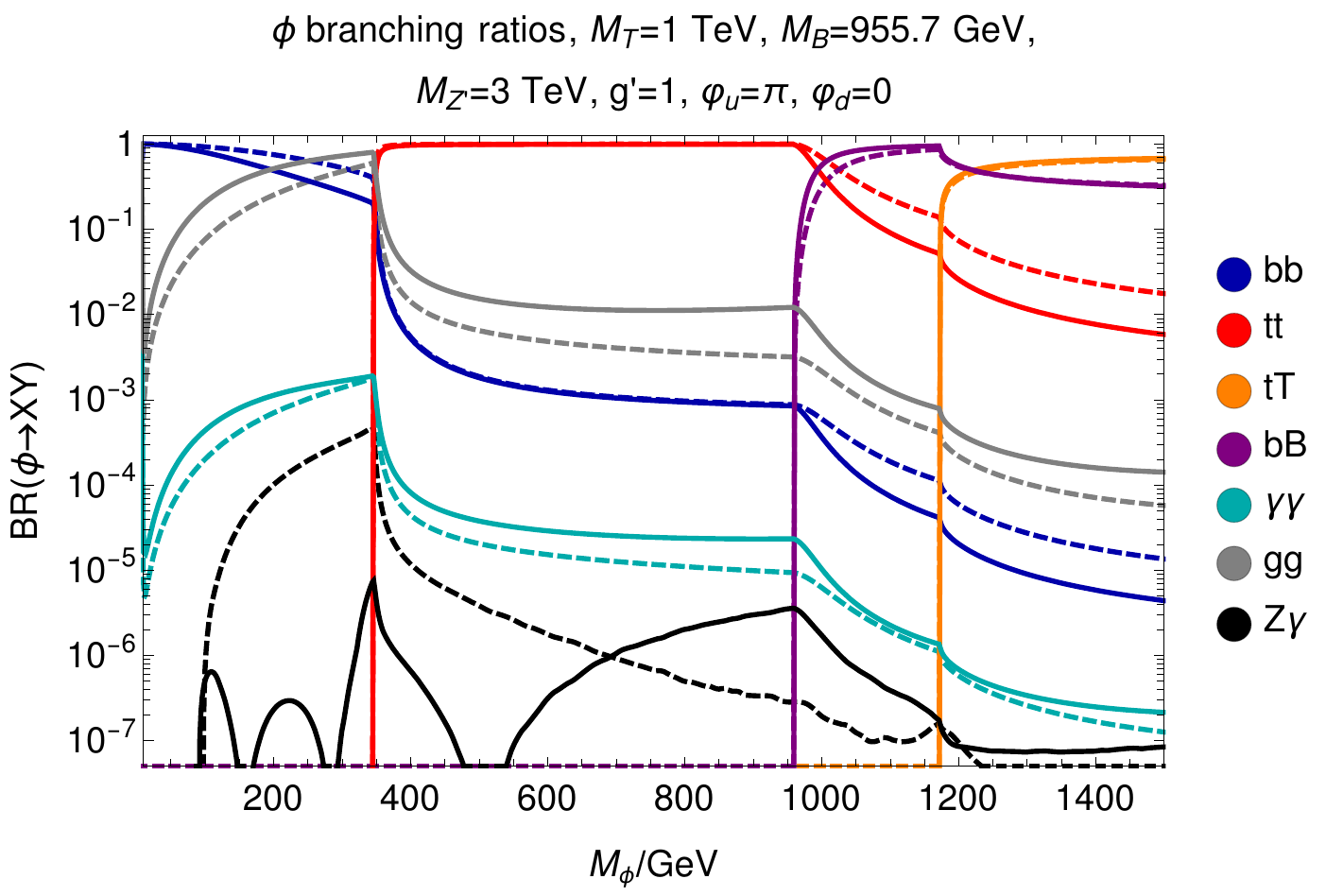}\\
	\caption{BRs of the Yukon $\phi$ as a function of $M_\phi$. Here, solid lines represent $\theta_Q^{34}=\pi/4$ whereas dashed ones denote $\theta_Q^{34}=\pi/6$. 
}
	\label{fig:phi_BRs}
\end{figure}

\section{Collider Signatures of the Higgs Singlet Yukon $\phi$}
\label{collider}
\subsection{Yukon $\phi$ Production at the LHC}
\label{sec:phi_production}

\begin{figure}[htbp]
	\centering
	\begin{subfigure}{0.3\textwidth}
		\includegraphics[scale=1.2]{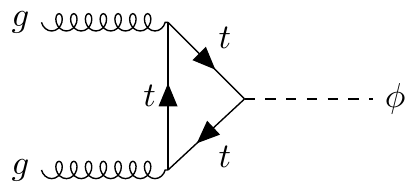}
	\end{subfigure}~~~~
	\begin{subfigure}{0.3\textwidth}
		\includegraphics[scale=1.2]{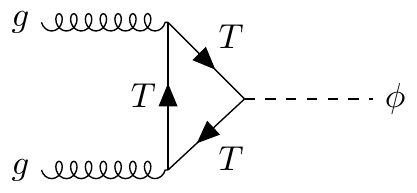}
	\end{subfigure}~~~~
\begin{subfigure}{0.3\textwidth}
	\includegraphics[scale=1.2]{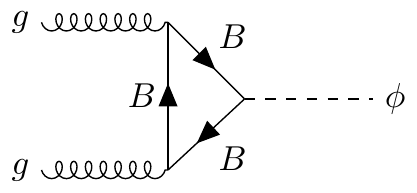}
\end{subfigure}
	\caption{Dominant Yukon $\phi$ production modes at the LHC.}
	\label{fig:phi_production_gluons}
\end{figure}

\begin{figure}[htbp]
	\centering
	\includegraphics[width=0.9\linewidth]{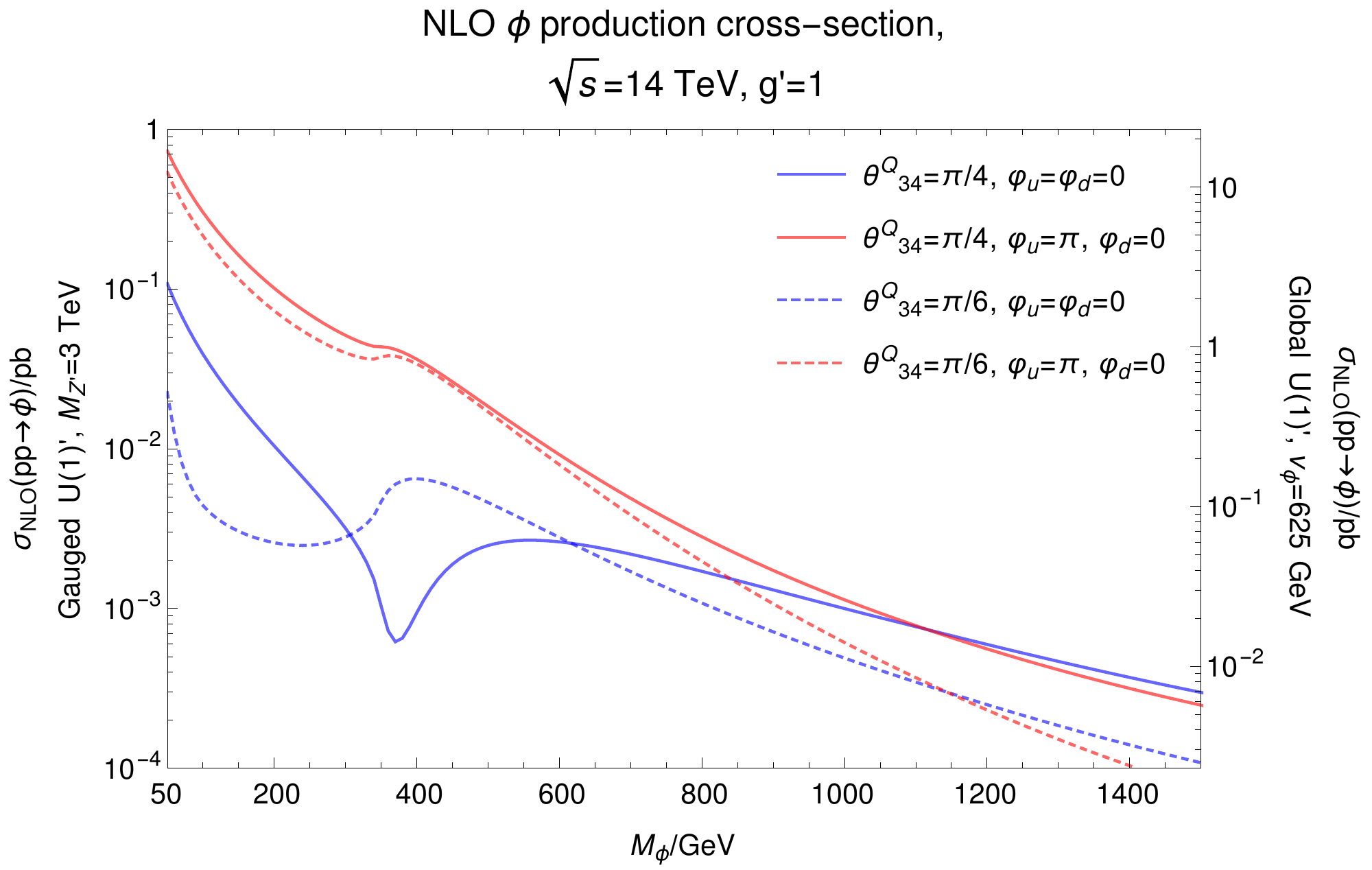}
	\caption{Production cross-section of the Yukon $\phi$ from ggF at the LHC with $\sqrt{s}=14$ TeV, via $t,T,b,B$ loops, with parameters: 
$\thQ = \pi /4$ in solid and $\thQ=\pi/6$ in dashed. In blue we plot for $(\varphi_u=\varphi_d=0)\simeq(\varphi_u=0, \varphi_d=\pi)$ and in red      
 for $(\varphi_u=\varphi_d=\pi)\simeq(\varphi_u=\pi, \varphi_d=0)$. The left scale is for the gauge model, $M_{Z'}=3$ TeV, and the right scale is for the global model, with $v_\phi=625$ GeV. Note that the shape of the curves for the gauge and global models is identical.}
	\label{fig:phi_production_with_phases_newzprregime_piover4}
\end{figure}

Using the same formulae as for the SM case, we calculate the production cross-section for the new scalar $\phi$ from ggF, as can be seen in the Feynman diagrams in Figure \ref{fig:phi_production_gluons}. This production cross-section is plotted with the change in $\phi$ mass in Figure \ref{fig:phi_production_with_phases_newzprregime_piover4}. We calculate this at NLO by employing a $k$-factor at NLO  of $1.7$, similarly to the SM Higgs production case. As previously, we fix the gauge coupling to be unity, $g'=1$, and take the EWPO and FCNC limits into account by using $M_{Z'}=3000$ GeV, in order to maximise the Yukawa couplings. In the solid colours we have taken $\thQ=\pi/4$ while in the dashed ones we have adopted $\thQ=\pi/6$. In addition, we plot two phase regimes, $\varphi_u=\varphi_d=0$ in blue and $\varphi_u=\pi,~\varphi_d=0$ in red. Modifying the down quark phase bears little impact due to the bottom quark mass suppresion (and so is not shown in the figure), but one can see that the $\varphi_u=\pi$ phase is clearly favourable due to interference effects as we now discuss. The $t$-loop contribution is subleading so that the kink around $M_\phi = 2 m_t$ is due to interference effects between the $t$-, $T$- and $B$-loop channels (in fact, we find that even for lower $\phi$ masses the VLT and VLB amplitudes still dominate), since the $\phi$ only couples to the SM-like top through mass mixing. The production cross-sections correspond to those of the gauged (global)  $U(1)'$ version of the model on the left(right)-hand side of the frame.

\begin{figure}[htbp]
	\centering
	\includegraphics[width=0.9\linewidth]{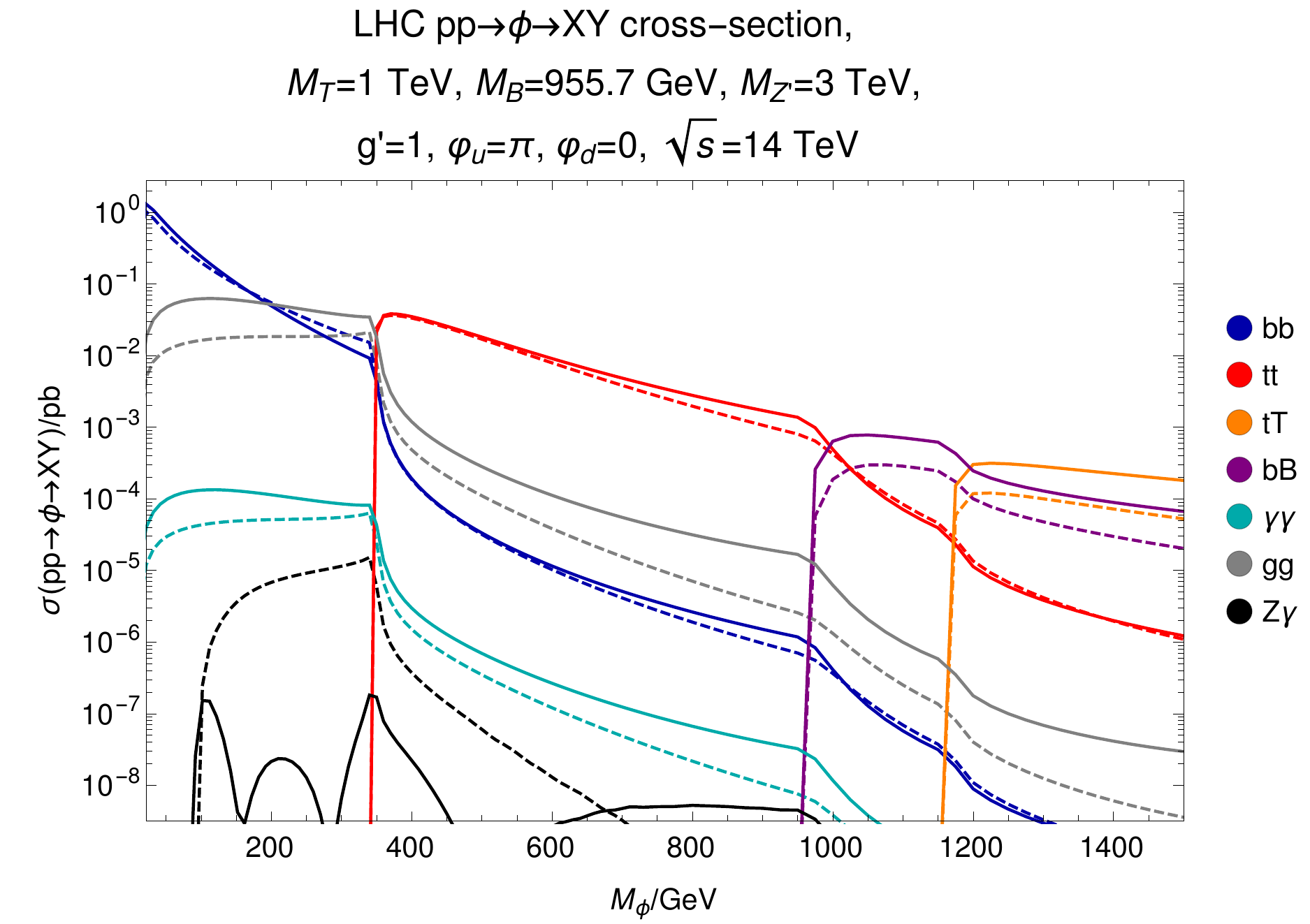}
	\caption{Rates for Higgs singlet Yukon $\phi$ production and decay at the LHC with $\sqrt{s}=14\text{TeV}$. Solid lines correspond to $\theta_{34}^Q=\pi/4$ and dashed lines to $\theta_{34}^Q=\pi/6$. Above 350 GeV the $tt$ mode dominates, suppressing the $\gamma \gamma$ signal, making the Yukon $\phi$ harder to discover.}
	\label{fig:phi_production_and_decay_at_the_LHC}
\end{figure}

\subsection{Discovery Channels for the Yukon $\phi$}
Now, we may collect together the results from previous sections to look at the full production and decay chain of the Yukon $\phi$. We utilise the set of parameters which maximises Yukawa couplings while complying with the EWPO/FCNC limits that have been discussed previously, along with the optimal transformation phases required to give a maximal production cross-section for $\phi$, as illustrated in section \ref{sec:phi_production}. The cross-section for $\phi$ production from ggF at NLO multiplied by the BR for each channel (computed at LO), with changing $\phi$ mass, is shown in Figure \ref{fig:phi_production_and_decay_at_the_LHC}, where we have included all possible kinematically-allowed decays of the new scalar. The cross-sections are much too small to compete with QCD backgrounds, so there is no chance to see a signal from $bb,~tt,~bB,~gg$ at any $M_\phi$. We will however study the possibility to discover the other channels, which suffer significantly less from QCD contamination, like $\phi$ $\to$ $tT$, $\gamma \gamma$ and $Z \gamma$, though we found that the latter is never competitive (hence we neglect it thereafter). 

We will consider two Yukon $\phi$ mass regimes, both of which are designed to offer maximum sensitivity to the two interesting channels i.e., $\gamma\gamma$ and $tT$. By investigating Figure \ref{fig:phi_production_and_decay_at_the_LHC}, two obvious choices emerge. We will study $\gamma\gamma$ for $M_\phi = 340$ GeV and $tT$ (with $T\to h^0 t,tZ$) for $M_\phi \approx 1250$ GeV. In all channels, we will consider both model configurations, the gauged as well as the global one. Then, as collider energies, we will use $\sqrt s =14$ TeV (which is appropriate for both Run 3 of the LHC and the HL-LHC) as well as $\sqrt s=33$ TeV (which is appropriate for both the HE-LHC and the first stage of the FCC). In all cases, we will use as integrated luminosity ($L^{int}$) the value of $3000$ fb${}^{-1}$, so as to ascertain the relative strength of each collider soilely in terms of energy reach. (In fact, we can anticipate that the model with a global $U(1)'$ symmetry offers some sensitivity in the $\gamma\gamma$ case already with 300 fb$^{-1}$ at the lower energy considered.)

To recap, for all three channels, we will use the following common input parameters:
\begin{gather}
	M_T=1000 \textrm{ GeV},~~~M_B=955\textrm{ GeV},~~~g'=1,~~~ \thQ=\pi/4,~~~ \varphi_u = \pi ,~~~ \varphi_d =0,\\
\begin{align}
	\textrm{Gauged model:}~ &M_{Z'}=v_\phi=3000 \textrm{ GeV}, \\
	\textrm{Global model:}~ &v_\phi=625 \textrm{ GeV},
	\end{align}
           \label{input}
	\end{gather}
and for specific signatures,
\begin{align}
\vspace*{-0.5cm}
	\gamma \gamma:&~~~ M_\phi=340 \textrm{ GeV}, \\
	T\to tth^0, ttZ:&~~~M_\phi=1250 \textrm{ GeV}.
	\end{align}
As intimated, these values correspond to the optimal sensitivity yield for all discovery channels considered. Specifically, note that the global model VEV of $v_\phi=625$ GeV is taken as the smallest possible one (to optimise Yukawa couplings) whilst still accounting for the perturbative limit set in Equation
(\ref{eq:vev_phi}) of $v_\phi \gtrsim M_\phi /2 = 625$ GeV.

\subsection{Methodology}
In all  channels, we will perform the usual ``bump hunt'' in the invariant mass plots by determining the Gaussian significance at the resonance of the $\phi$ state, for a given mass $M_\phi$. For both signal and background, we will calculate the cross-sections at LO and approximate the NLO result by employing the relevant $k$-factors. We calculate the signal cross-section as previously described while using \texttt{MadGraph} \cite{Stelzer:1994ta} to calculate the backgrounds \footnote{Throughout this work we have used version \texttt{MG5\_aMC\_v2.6.0}, with default parameters unless otherwise specified.}. For each channel we will identify the detector resolution and find the signal and background cross-section within the relevant invariant mass window. To account for the number of observed events by the detector, we will employ an acceptance $\times$ selection efficiency factor to both signal and background. We find the Gaussian significance of the resonance by the usual formula:
\begin{equation}
	\alpha = \frac{S}{\sqrt{S+B}},
	\end{equation}
where $S$ is the number of signal events and $B$ is the value for the background ones after they have undergone a kinematical selection with an
acceptance $\times$ selection efficiency rate $\epsilon(S)$ and $\epsilon(B)$, respectively.

For later convenience we summarise the cross-sections and cuts which we shall assume in the remainder of the paper.
The cross-sections for Yukon $\phi$ production and decay in various channels are given in Table~\ref{tab:Xsecs-BRs}.
The cross-sections and cuts on SM background processes in various channels are shown in Table~\ref{tab:all_info}.
These numbers will be used to calculate the final significances obtained in the later results
in Table \ref{tab:gamma_gamma_discovery_significances}--\ref{tab:ttZ_discovery_significances}.

\begin{table}[thbp]
	\centering
			\resizebox{1.\textwidth}{!}{%
		\begin{tabular}{|c|c|c|c|c|c|c|c|c|} \hline		
			\multirow{2}{*}{Channel} &\multirow{2}{*}{Energy} & \multirow{2}{*}{$M_\phi$ (GeV)} & \multicolumn{2}{c|}{$\sigma_{\rm NLO}(pp \rightarrow \phi)$  (pb)}&  \multirow{2}{*}{Branching Ratio} & \multirow{2}{*}{Cuts} & \multicolumn{2}{c|}{Final Cross-Section (pb)} \\ \cline{4-5} \cline{8-9}
			&&&Gauge&Global&&&Gauge&Global\\ \hline
			\multirow{2}{*}{$\gamma \gamma$}&$\sqrt{s}=14$ TeV & \multirow{2}{*}{340} & 0.0437&1.01 & \multirow{2}{*}{BR($\phi \rightarrow \gamma \gamma)=0.00186$} & \multirow{2}{*}{$\epsilon=0.81 ^*$} & $6.60 \times  10^{-5}$&0.0015 \\
			&$\sqrt{s}=33$ TeV &   & 0.228 &5.25& &&   0.00034&0.0078 \\	\hline
			 \multirow{2}{*}{$tth^0$}&{$\sqrt{s}=14$ TeV} & \multirow{2}{*}{1250} & {0.000478} &0.0110& \multirow{2}{*}{BR($\phi \rightarrow tT \rightarrow tth^0)=0.378$} & \multirow{2}{*}{$\epsilon(S)$} & $0.00018 ~ \epsilon(S)$&$0.00441 ~ \epsilon(S)$ \\
 			&{$\sqrt{s}=33$ TeV} &  &{0.00594}&0.137&  &  & $0.0022 ~ \epsilon(S) $&$0.051 ~ \epsilon(S) $  \\ 	\hline
 			 \multirow{2}{*}{$ttZ$}&{$\sqrt{s}=14$ TeV}&\multirow{2}{*}{1250} & {0.000478}&0.0110& \multirow{2}{*}{BR($\phi \rightarrow tT \rightarrow ttZ)=0.242$} & \multirow{2}{*}{$\epsilon(S)$}& $0.00012  ~\epsilon(S)$&$0.0028  ~\epsilon(S) $ \\
 			&{$\sqrt{s}=14$ TeV}&&{0.00594}&0.137&&& $0.0014  ~\epsilon(S)$&$0.032  ~\epsilon(S) $  \\ \hline
		\end{tabular}%
	}
	\caption{Table of cross-sections for Yukon $\phi$ production and decay in various channels, with 
	cuts on the signal processes. Note all signals are calculated at LO, then multiplied by a 
$k$-factor of $k^{\rm NLO}=1.7$ to get the written NLO results: $S^{\rm NLO}=S^{\rm LO} \times k^{\rm NLO}$. 
		The parameter set used in all cases is as follows: $\{ \thQ=\pi/4 ,~ \varphi_u =0,~\varphi_d=\pi,~g'=1\}$. Gauged model fixes $M_{Z'}=3$ TeV, whereas global model fixes $v_\phi=625$ GeV.
		 Results for $v_\phi = 625$ are a factor $(3000/625)^2 \simeq 23$ larger than the $M_{Z'}=3$ TeV ones in all cases, since $\sigma_{\rm NLO}(pp \rightarrow \phi) \propto v_\phi ^{-2}$ and BRs are independent of $v_\phi$ in all channels.\\
		${}^{*}$Signal cut effect on $\phi \rightarrow \gamma \gamma$ determined from fraction of SM $h^0 \rightarrow \gamma \gamma$ events observed with and without $\eta,~p_T$ cuts. }
\label{tab:Xsecs-BRs}
\vspace{1cm}

	\centering
	\resizebox{1.\textwidth}{!}{%
	\begin{tabular}{|c|c|c|c|c|c|}	\hline	
Channel &  Energy & Cuts (GeV, except $\eta$) & $\sigma_{LO} (pp \rightarrow X)$ (pb) & $k$-Factor & Final Cross-Section (pb) \\ \hline
\multirow{2}{*}{$\gamma \gamma$} & $\sqrt{s}=14$ TeV & \multirow{2}{*}{$\bigg \{$} $|\eta| < 2.5,~p_T > 25$ GeV  \multirow{2}{*}{$\bigg \}$}& 0.0157 & 1.88 &  0.0295 \\
 & $\sqrt{s}=33$ TeV & $336<M_{\gamma\gamma}<344$ & 0.0328 & 1.88 &  0.0617 \\ \hline
 \multirow{2}{*}{$tth^0$} & $\sqrt{s}=14$ TeV & \multirow{2}{*}{$\bigg \{$} $850 < M_{h^0 t}<1150 $ \multirow{2}{*}{$\bigg \}$} &  0.00603${}^\dagger$ & 1.27 &  $0.0153\times \epsilon(B)$ \\
 &$\sqrt{s}=33$ TeV & $1063<M_{t{t}h^0}<1438$ & 0.0542${}^\dagger$ & 1.27 & $0.138 \times \epsilon(B)$ \\ \hline
  \multirow{2}{*}{$ttZ$}& $\sqrt{s}=14$ TeV &  \multirow{2}{*}{$\bigg \{$} $900<M_{tZ}<1100$ \multirow{2}{*}{$\bigg \}$} &  0.0064${}^\dagger$ & 1.43 &  $0.0183 \times \epsilon(B)$\\
  & $\sqrt{s}=33$ TeV & $1125<M_{ttZ}<1375$ & 0.0579${}^\dagger$ & 1.43 & $0.166  \times \epsilon(B)$ \\ \hline
	\end{tabular}%
}
	\caption{Table of cross-sections and cuts on SM background processes in various channels, which will compete against the $\phi$ boson signal in these channels shown in the previous table, where the suggested cuts are designed to enhance the signal.\\
		$^\dagger$The listed $\sigma_{LO}$ results are calculated using a cut on $M_{h^0 t}$ (or $M_{tZ}$). To account for the alternative cut, on $M_{\bar{t}h^0}$ (or $M_{\bar{t}Z}$), one should multiply the $\sigma _{LO}$ result by a factor of 2, which is included in the final cross-section.}
	\label{tab:all_info}
\end{table}

\subsection{The $\gamma \gamma$ signal}

\begin{figure}[h]
	\centering
	\begin{subfigure}{0.3\textwidth}
		\includegraphics[scale=1.2]{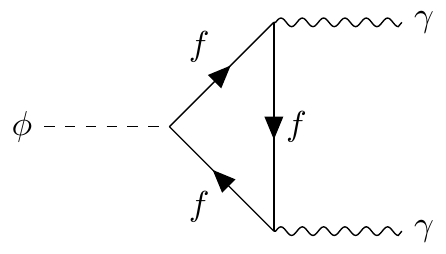}
	\end{subfigure}
	\caption{Feynman diagram for $\phi$ $\to$ $\gamma \gamma$, where $f=\{b,~t,~B,~T\}$.}
	\label{fig:phi_decay_gamma_gamma}
\end{figure}

\begin{table}[h]
	\centering
	\begin{tabular}{|c|c|c|}
		\hline
		Model & Experiment ($L^\textrm{int}=3000~\text{fb}^{-1}$) & Significance \\ \hline
		Gauged $U(1)'$ &HL-LHC, $\sqrt{s}=14$ TeV & 0.66$\sigma$ \\
		$M_{Z'}=3000$ GeV&HE-LHC/FCC, $\sqrt{s}=33$ TeV & 2.4$\sigma$ \\ \hline
		Global $U(1)'$ & 
		{HL-LHC, $\sqrt{s}=14$ TeV} & {15$\sigma$} \\		 
		$v_\phi=625$ GeV &HE-LHC/FCC, $\sqrt{s}=33$ TeV & 52$\sigma$\\
		\hline		
	\end{tabular}
	\caption{Significances for the $gg\to \phi\to\gamma\gamma$ signal with  $M_\phi = 340$ GeV and the parameter setup given in Equation~(\ref{input}) after the following cuts on both photons: $|\eta| < 2.5$ and $p_T > 25$ GeV.}
	\label{tab:gamma_gamma_discovery_significances}
\end{table}

We will now examine the possibility to detect the Yukon $\phi$ through ggF, as seen in the Feynman diagrams in Figure \ref{fig:phi_production_gluons}, and decaying to $\gamma \gamma$, as shown in the Feynman diagram in Figure \ref{fig:phi_decay_gamma_gamma}. As intimated, at low Yukon masses, namely for $M_\phi < 2 m_t$, this is the cleanest and simplest channel to consider. The relevant cuts to adopt are on the pseudorapidity and transverse momentum of the photons, both taken with $|\eta |<2.5$ and $p_T > 25$. To simulate the effect of these cuts on our signal, we calculate the fraction of events captured by these through a Monte Carlo (MC) simulation and obtain an acceptance $\times$ selection efficiency rate of $\epsilon(S)=0.81$. For the two collider configurations considered (HL-LHC and HE-LHC/FCC), we assume a detector resolution of $2.5\%$ \cite{1804.02716}.  

Taking 340 GeV for the Yukon mass, this   leads to an invariant mass window on the photons of $336$ GeV $<M_{\gamma \gamma}<344$ GeV, over which we sample both signal and background. The latter, like the former,  is also generated at LO, yet supplemented by a $M_{\gamma\gamma}$  dependent NLO $k$-factor obtained from \cite{Catani:2018krb} (which, at $M_{\gamma \gamma}=340$ GeV, gives  $1.88$) with, again computed through MC analysis,  of $\epsilon(B)\simeq 1$ \footnote{The efficiency is around 0.96 from \cite{Aad:2019wsl}, but within the accuracy of this paper we approximate this to unity.}. 

Based on the above kinematical selection, we find the significances given in Table \ref{tab:gamma_gamma_discovery_significances}. From here, we can see that the gauged model would be difficult to find at the HL-LHC, whereas there could be some indication of this signal at the
HE-LHC and FCC. The global model, however, can easily be seen in all such collider environments. Indeed, the latter also has clear potential to be accessed by the end of Run 3 of the LHC, assuming  $L^\textrm{int}=300~\textrm{fb}^{-1}$, as the significance rescales to a $4.7\sigma$ signal. All this is modulo the effects of photon identification and of mistagging jets and/or electrons as photons, both of which are however expected to be marginal. Finally, notice that we refrain here from placing  exclusion limits on our model, as this is complicated by the dimensionality of the parameter space, which is mapped in $\{M_\phi,~ \thQ,~ g',\varphi_u,~\varphi_d,~v_\phi,~M_T,~M_B\}$. Indeed, we leave this task for future studies.

\subsection{The $t {t} h^0$ Signal}
\label{sec:tth_signal}

\begin{figure}[h]
	\centering
	\begin{subfigure}{0.3\textwidth}
		\includegraphics[scale=1.2]{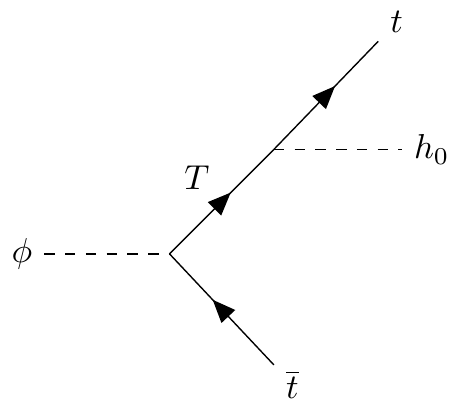}
		\end{subfigure}
\caption{Feynman diagram for Yukon decay $\phi \rightarrow t{t}h^0$.}
\label{fig:feyn_tth}
\end{figure}

\begin{table}[h]
	\centering
	\begin{tabular}{|c|c|c|c||c|c|}
		\hline
		\multirow{4}{*}{Model} & \multirow{4}{*}{Experiment ($L^\textrm{int}=3000~\text{fb}^{-1})$} & \multicolumn{4}{c|}{Significance} \\ \cline{3-6}
		&&\multicolumn{2}{c||}{$\epsilon(S)=\epsilon(B)$} & \multicolumn{2}{c|}{$\epsilon(S)=2\epsilon(B)$}\\ \cline{3-6}
		&&\multicolumn{2}{c||}{$\epsilon(B) \cdot 10^{-3}$} & \multicolumn{2}{c|}{$\epsilon(B) \cdot 10^{-3}$} \\ \cline{3-6}
		&& 5&10 & 5&10 \\ \hline
		Gauged $U(1)'$ &HL-LHC, $\sqrt{s}=14$ TeV & 0.056$\sigma$ & 0.080$\sigma$ & 0.11$\sigma$ &0.16$\sigma$\\
		$M_{Z'}=3000$ GeV&HE-LHC/FCC, $\sqrt{s}=33$ TeV &0.23$\sigma$ & 0.33$\sigma$ & 0.46$\sigma$ &0.65$\sigma$ \\ \hline
		Global $U(1)'$ & {HL-LHC, $\sqrt{s}=14$ TeV} & {1.2$\sigma$}&{1.6$\sigma$}&{2.1$\sigma$}&{3.0$\sigma$}\\
		$v_\phi=625$ GeV&HE-LHC/FCC, $\sqrt{s}=33$ TeV &4.6$\sigma$ & 8.2$\sigma$ & 6.5$\sigma$ &12$\sigma$ \\ \hline		
	\end{tabular}
	\caption{Significances for the $gg\to \phi\to tT\to tth^0$ signal with  $M_\phi = 1250$ GeV and the parameter setup given in Equation~(\ref{input}) for four values of acceptance $\times$ selection efficiency: on the left the total signal efficiency is equal to the background one, $\epsilon(S)=\epsilon(B)$,
		while on the right it is twice the background one, $\epsilon(S)=2 \epsilon(B)$. For each of those cases, we take two background total efficiencies: a conservative $\epsilon(B)=5 \times 10^{-3}$ and an optimistic $\epsilon(B)= 10^{-4}$. 
	}
	\label{tab:tth_discovery_significances}
\end{table}

In addition to the previous mode, there is also the possibility to detect $gg\to \phi\rightarrow tT\rightarrow tth^0$, as shown in Figure \ref{fig:feyn_tth}, when the new scalar state is heavy. To gain sensitivity in this channel (specifically, to the $M_\phi$ resonance), we can exploit the  decay of the VLT, as the invariant mass of its decay products will be equal to its mass, $M_{h^0 t}=M_T$, and in turn we will also have $M_{tth^0}=M_\phi$, up to some detector resolution. 
Unlike  the $\gamma \gamma$ case, $t{t} h^0$ is reconstructed through several different channels which depend on the various possible decay paths of the top and Higgs states. Furthermore,
these $tT$ channels are very complicated to reconstruct and suffer
from important backgrounds like $t\bar t$  and others that are extremely large.
This leads to a far worse $M_{tth^0}$  invariant mass resolution compared to the $M_{\gamma\gamma}$ case, of some 30\% \cite{private_communication}, which we will also use as the resolution for the invariant mass of the VLT decay products, $M_{h^0 t}$. 
The adoption of such a predefined value of resolution, rather that using 
 physical objects like those appearing in detectors and emulated through MC event generation, is clearly 
 far from  real experimental analyses, so that we are bound to produce only rough estimates of the sensitivity to this process at present and future hadron colliders. Nonetheless, with this approach, we aim at spurring more sophisticated phenomenological analyses.

Currently, at the LHC with $\sqrt{s}=13$ TeV,  a total cross-section of $\sigma(t t h^0)=790 ^{+230} _{-210}$ fb from ATLAS \cite{Aaboud:2017jvq} and $\sigma(t {t} h^0)=639 ^{+157} _{-130}$ fb from CMS\footnote{What is reported from CMS is the signal normalised to the SM prediction of 1.26 $^{+0.31} _{-0.26}$, from which we can extract the above cross-section.} has been measured \cite{1804.02610}. 
However, for the luminosity recorded to date at Run 2, the event rates are too small to plot any meaningful invariant mass distribution in this channel\footnote{We also note that such $t\bar t h^0$  data disagree with SM predictions, to the extent that, presently, it is unclear whether one is in presence of genuine anomalies due to BSM physics or merely an inadequate estimate of the SM processes.}, so we have to compute these by MC analysis, which we do again with {\tt MadGraph}. Here, we employ a constant $k$-factor to calculate the NLO corrections from QCD for the background of this channel as $k_{tth^0}=1.27$, derived from comparing the LO \texttt{MadGraph} result to the SM prediction of $\sigma (t {t}h^0)=507 ^{+35} _{-50}$ fb given in \cite{Aaboud:2017jvq}. We then compute significances using a conservative total efficiency (fraction of events which are recorded after all cuts have been applied) of $\epsilon(B)=0.005$ from \cite{Aaboud:2017jvq} and also display the significances for a more optimistic scenario where this is doubled to $\epsilon(B)=0.01$. This scenario represents a potential total efficiency when incorporating all possible decay paths, including $h^0 \rightarrow \bar{b}b$, which was not incorporated in the above study. We also assume two scenarios for the signal total efficiency: a conservative one, where it is equal to the background case $\epsilon(S)=\epsilon(B)$, and another optimistic one, where the total efficiency is twice that of the background, $\epsilon(S)=2\epsilon(B)$. We present the yields of these scenarios in Table \ref{tab:tth_discovery_significances}, wherein it should be recalled that we are assuming $30\%$ resolution for both $M_{h^0 t}$ and $M_{tth^0}$. With this resolution, and having fixed $M_\phi$ just above the $tT$ threshold at 1250 GeV, the invariant mass cuts that we apply are $850$ GeV $<M_{h^0 t}<1150$ GeV and $1063$ GeV $<M_{tth^0}<1438$ GeV. We finally find that the significance is too small for the signal to  be seen at the HL-LHC, or even at the HE-LHC/FCC, for the gauged model, though the global one will show a clear signal already at the HL-LHC with plenty of discovery potential at the HE-LHC/FCC.

\subsection{The $t {t} Z$ Signal}

\begin{figure}[h]
	\centering
	\includegraphics[scale=1.2]{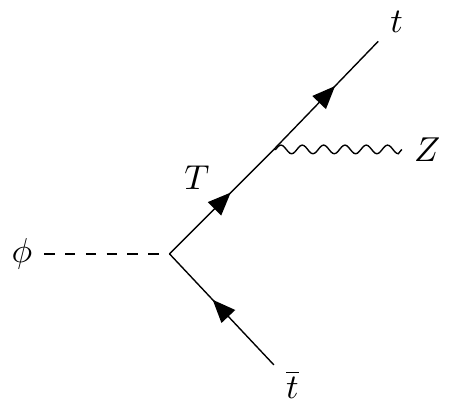}
	\caption{Feynman diagram for Yukon decay $\phi \rightarrow t{t}Z$.}
	\label{fig:feyn_ttZ}
\end{figure}

\begin{table}[h]
	\centering
	\begin{tabular}{|c|c|c|c||c|c|}
		\hline
		\multirow{4}{*}{Model} & \multirow{4}{*}{Experiment ($L^\textrm{int}=3000~\text{fb}^{-1})$} & \multicolumn{4}{c|}{Significance} \\ \cline{3-6}
		&&\multicolumn{2}{c||}{$\epsilon(S)=\epsilon(B)$} & \multicolumn{2}{c|}{$\epsilon(S)=2\epsilon(B)$}\\ \cline{3-6}
		&&\multicolumn{2}{c||}{$\epsilon(B) \cdot 10^{-3}$} & \multicolumn{2}{c|}{$\epsilon(B) \cdot 10^{-3}$} \\ \cline{3-6}
		&& 5&10 & 5&10 \\ \hline
		Gauged $U(1)'$ &HL-LHC, $\sqrt{s}=14$ TeV & 0.033$\sigma$ & 0.047$\sigma$ & 0.066$\sigma$ &0.093$\sigma$\\
		$M_{Z'}=3000$ GeV&HE-LHC/FCC, $\sqrt{s}=33$ TeV & 0.14$\sigma$ & 0.19$\sigma$ & 0.27$\sigma$ &0.38$\sigma$ \\ \hline
		Global $U(1)'$ & {HL-LHC, $\sqrt{s}=14$ TeV} & {0.71$\sigma$}&{1.0$\sigma$}&{1.3$\sigma$}&{1.9$\sigma$}\\
		$v_\phi=625$ GeV&HE-LHC/FCC, $\sqrt{s}=33$ TeV &2.9$\sigma$ & 4.1$\sigma$ & 5.3$\sigma$ &7.5$\sigma$ \\ \hline			
	\end{tabular}
	\caption{Significances for the $gg\to \phi\to tT\to ttZ$ signal with  $M_\phi = 1250$ GeV and the parameter setup given in Equation~(\ref{input}) for four values of acceptance $\times$ selection efficiency: on the left the total signal efficiency is equal to the background one, $\epsilon(S)=\epsilon(B)$,
		while on the right it is twice the background one, $\epsilon(S)=2 \epsilon(B)$. For each of those cases, we take two background total efficiencies: a conservative $\epsilon(B)=5 \times 10^{-3}$ and an optimistic $\epsilon(B)= 10^{-4}$. 
	}
	\label{tab:ttZ_discovery_significances}
\end{table}

We may proceed with $t {t} Z$ in a similar fashion to $t {t} h^0$\footnote{Hence, with a similar caveat regarding the accuracy of our predictions.}, utilising now the decay path $T \rightarrow Zt$ as shown in Figure \ref{fig:feyn_ttZ}. With respect to the $tth^0$ case, though. we may assume here a better resolution of $20\%$\footnote{Owing to the fact that a significant portion of $Z$ decays is into electrons and muons, while the SM Higgs state essentially only decays into $b\bar b$ and $W^+W^-\to 4$-fermions, i.e., predominantly into hadronic final states, which are more difficult to reconstruct in comparison.},  
for both  $M_{tZ}$ and $M_{t{t}Z}$. Hence, we can adopt the invariant mass cuts   $900$ GeV $<M_{tZ}<1100$ GeV and $1125$ GeV $<M_{ttZ}<1375$ GeV, as we again have  $M_\phi=1250$ GeV. We employ as NLO $k$-factor for the background the value $k_{ttZ}=1.43$, which is constant in our MC generation. We list the results for the same total efficiencies as in the $t{t}h^0$ case,  $\epsilon(S)=\epsilon(B)$ and $\epsilon(S)=2\epsilon(B)$, with $\epsilon(B)=5 \times 10^{-3}$ and $\epsilon(B)=10^{-4}$,  in Table \ref{tab:ttZ_discovery_significances}. The significances are similar to the $t {t} h^0$ signature, so that the gauged model will remain difficult to trace anywhere in the large $\phi$ mass scenario. Similarly to $tth^0$, the $ttZ$ channel offers a small signal for the global model at the HL-LHC and significant discovery potential at the HE-LHC/FCC.

\section{Conclusions}
\label{conclusions}

Amongst the many puzzles pertaining to the structure of the SM is the origin of the Yukawa couplings, 
responsible for the strong hierarchy of the quark and lepton masses. 
A possible explanation for this feature could be their different origin, for example, rather than emerging as direct couplings to the SM Higgs doublet, effective Yukawa couplings 
could be generated via their mixing with new vector-like fermionic states hitherto undiscovered. 
An intriguing possibility in this respect could be the one realised through an additional, gauged or global, $U(1)'$ symmetry added to the SM gauge group, which explicitly forbids all direct Yukawa couplings at the Lagrangian level. Such couplings are effectively generated after 
$U(1)'$ breaking, via seesaw type diagrams, involving both Higgs doublets and a new Higgs singlet, where the strength of the Yukawa coupling is suppressed by the mass of the intermediate vector-like fermion.

The large third family quark $(t,b)$ Yukawa couplings are effectively generated via mixing with a vector-like fourth family quark EW doublet 
$(T,B)$, which are assumed to be relatively light, with masses perhaps at the TeV scale. 
The smallness of the second family quark $(c,s)$ Yukawa couplings is due to their coupling to heavier vector-like fourth family quark EW singlets, whose masses must lie well beyond current collider energies. Similar considerations apply to the lightest first family quarks $(u,d)$ which couple to even heavier VLQs. The hierarchy of leptonic Yukawa couplings can similarly result from vector-like leptons, although these are more difficult to produce at the LHC.

Hence in this paper we have focussed on a simplified model involving only 
$(t,b)$ and the lightest VLQ EW doublets 
$(T,B)$, whose masses may lie around the TeV scale where they would be accessible to the LHC.
Furthermore, following the $U(1)'$ breaking, a new Higgs singlet state, $\phi$, which we refer to as the Yukon,
is generated, with characteristic couplings to both $(t,b)$ and $(T,B)$ quarks. If the 125 GeV Higgs boson $h^0$ is ``the origin of mass'', then the Yukon $\phi$ is ``the origin of Yukawa couplings'', and its discovery would be a smoking gun of this mechanism.
To be more precise, we have seen that what distinguishes this model from the SM plus VLQs is the existence of the Yukon with couplings to fermions proportional to the fermion masses, a feature shared by that of the SM Higgs boson.

In the considered simplified model, the masses of both the new fermion doublet  $(T,B)$ and Higgs singlet Yukon $\phi$ are essentially free parameters, so that one can attempt to access their signals at present and/or future hadron colliders, such as the LHC (Run 3), HL-LHC, HE-LHC and/or FCC. Presently, the lower bound on their masses, $M_T\sim M_B$, is constrained to be at the TeV scale or just below it, by  direct searches at Run 2 of the LHC, as such vector-like (coloured) states can be copiously produced via standard QCD interactions. On the other hand, the constraints on the new Higgs singlet $\phi$ mass $M_\phi$ are very weak, as the new scalar state $\phi$ does not  couple directly to any SM particle but the $t$ and $b$ quarks, whose couplings remain small in comparison to those involving $T$ and $B$ states. Therefore, a natural way of establishing some sensitivity to this model is to exploit signatures where all such new states, $T,B$ and $\phi$, coexist. Specifically, one can look at the time-honoured production mode of a Higgs scalar at hadronic machines, i.e., gluon-gluon fusion (ggF), wherein the $\phi$ state is produced primarily via the $T$ and $B$ loops. Furthermore, depending on the relative values of $M_\phi$ and $M_T\sim M_B$, one can search for $\phi\to \gamma\gamma$ signals if $\phi$ is lighter than $T$ and $B$ (so that loops of the latter also trigger the di-photon decay of the former) or $\phi \to tT\to tth^0,ttZ$ otherwise (so that VLQs can instead be produced as real particles)\footnote{We expect the $\phi\to bB$ modes to be more difficult to extract owing to a significant SM backgrounds from $b$ quarks.}. 

For the purpose of extracting such hallmark signatures of our model, following an initial investigation of its parameter space compliant with both theoretical and experimental constraints, we have defined within it two benchmark points  which maximise the sensitivity of the aforementioned hadronic machines to the processes $gg\to \phi\to \gamma\gamma$ and $gg\to \phi\to tT\to tth^0,ttZ$ in the presence of the corresponding irreducible backgrounds, for the choices
$M_\phi=340$ and 1250 GeV, respectively, while setting $M_T\sim M_B\sim1$ TeV (i.e., just beyond the current limits). Upon performing a MC analysis of both, in the presence of NLO effects, we have been able to show that the di-photon signal could already by accessed during Run 3 of the LHC (at  nearly $5\sigma$ level), with the HL-LHC, HE-LHC and FCC providing incontrovertible evidence of it, albeit limitedly to the case of the global $U(1)'$ model.  As for the signal yielding $h^0,Z$ production in association with top-quark pairs, this will be very difficult to establish at the HL-LHC, though a combination of the individual significances corresponding the two cases ($h^0$ and $Z$) could produce herein an excess very close to the $5\sigma$ level, with the HE-LHC/FCC again offering plenty of scope for discovery. However, it continues to be the case that this is true only for the  global $U(1)'$ configuration, with the gauged $U(1)'$ one remaining very elusive.  We should also mention that the $tth^0,ttZ$ channels are quite involved and warrant a more sophisticated analysis than that considered in this paper, where we neglect important backgrounds such as $tt$ and do not decay the final state particles.

In conclusion, the discovery of the Higgs singlet Yukon $\phi$, with the predicted couplings to $(t,b)$ and $(T,B)$ in the simplified global $U(1)'$ model, would provide evidence for a new theory of Yukawa couplings. Our results indicate that the production cross-section of the Yukon with a mass around 300-350 GeV 
could be sufficiently enhanced by the vector-like quark doublet $(T,B)$ in the loop to enable its di-photon decays to be observed at the LHC Run 3. However, the observation of a Yukon with a mass above the TeV scale, decaying in the $tth^0$ and/or $ttZ$ channels, would require a future collider such as the HE-LHC or FCC.

\subsection*{Acknowledgements}

S.\,F.\,K. and S. M. acknowledge the STFC Consolidated Grant ST/L000296/1. S. F. K.  also acknowledges the European Union's Horizon 2020 Research and Innovation programme under Marie Sk\l{}odowska-Curie grant agreement
HIDDeN European ITN project (H2020-MSCA-ITN-2019//860881-HIDDeN).
S. M. is supported in part through the NExT Institute. The authors thank Billy Ford and Claire Shepherd-Themistocleous for useful discussions.

\appendix

\section{Two Higgs Doublets with the Higgs singlet}
\label{higgs}
The Higgs doublet and singlet VEVs are defined by

\begin{eqnarray}
& H_u = 
\begin{pmatrix}
H_u^+ \\ 
v_u + \frac{1}{\sqrt{2}}\left( \func{Re}H_{u}^0 + i\func{Im}H_{u}^0 \right)%
\end{pmatrix}%
, \quad H_d = 
\begin{pmatrix}
v_d + \frac{1}{\sqrt{2}}\left( \func{Re}H_{d}^0 + i\func{Im}H_{d}^0 \right)
, \quad H_d^-%
\end{pmatrix}%
\\
& \Phi = \frac{1}{\sqrt{2}} \left( v_\phi + \phi + i\func{Im}%
\Phi \right).  \label{eqn:Higgs_fields_mass_basis}
\end{eqnarray}%

\subsection{The potential for the 2HDM with the Higgs singlet}
As recently discussed in \cite{Hernandez:2021tii},
the scalar potential of the model under consideration takes the form:
\begin{align}
\begin{split}
V &=\mu _{1}^{2}H_{u}H_{u}^{\dagger } +\mu _{2}^{2}
H_{d}H_{d}^{\dagger } +\mu _{3}^{2}\Phi \Phi ^{\ast }
-\mu _{sb}^{2}(\Phi ^{2}+\left( \Phi ^{\ast }\right) ^{2})
+\lambda _{1}(H_{u}H_{u}^{\dagger }) ^{2}+\lambda _{2}
(H_{d}H_{d}^{\dagger })^{2} \\
&+\lambda _{3}(H_{u}H_{u}^{\dagger })(H_{d}H_{d}^{\dagger }) +\lambda _{4}(H_{u}H_{d}^{\dagger})(H_{d}H_{u}^{\dagger }) 
+\lambda _{5}(\varepsilon
_{ij}H_{u}^{i}H_{d}^{j}\Phi ^{2}+H.c.) \\
&+\lambda _{6}(\Phi \Phi ^{\ast })^2
+\lambda _{7}(\Phi\Phi ^{\ast })(H_{u}H_{u}^{\dagger }) 
+\lambda _{8}(\Phi \Phi ^{\ast })(H_{d}H_{d}^{\dagger })
\label{scalarpotential} 
\end{split}
\end{align}
where the $\lambda _{i}$ are dimensionless and $\mu _{j}$ ($j=1,2,3$) are dimensionful parameters.
Note that the $\mu _{sb}^{2}$ is a mass squared parameter which we include only for the global $U(1)'$ model,
where it softly breaks the $U(1)'$ symmetry of the Higgs potential,
giving non-zero mass to the would-be Goldstone boson resulting from the CP-odd mass matrix.
For large values of $\mu _{sb}$ the mass of the CP-odd would-be Goldstone boson becomes very large and 
is not relevant for phenomenology. However for smaller values of $\mu _{sb}$ this state would mix with the other CP-odd states 
and would have interesting phenomenological implications, which however we do not study in this paper.
In the gauged $U(1)'$ model, the $\mu _{sb}^{2}$ term is absent and the Goldstone boson is eaten by the $Z'$.

\subsection{Mass matrix for CP-even, CP-odd neutral and charged scalars}

\label{subsec:mass_mamtrix_for_CP_even_odd_charged} The squared mass matrix
for the CP-even scalars in the basis $\left( \func{Re}H_{u}^{0},\func{Re}%
H_{d}^{0},\phi \right) $ takes the form:%
\begin{equation}
{M}_{CP-\text{even}}^{2}=\left( 
\begin{array}{ccc}
4\lambda _{1}v_{u}^{2}-\frac{\lambda _{5}v_{u}v_{\phi}^{2}}{2v_{u}} & \frac{1}{2%
}\lambda _{5}v_{\phi}^{2}+2\lambda _{3}v_{u}v_{d} & \sqrt{2}v_{\phi}\left( \lambda
_{5}v_{d}+\lambda _{7}v_{u}\right) \\ 
\frac{1}{2}\lambda _{5}v_{\phi}^{2}+2\lambda _{3}v_{u}v_{d} & 4\lambda
_{2}v_{d}^{2}-\frac{\lambda _{5}v_{u}v_{\phi}^{2}}{2v_{d}} & \sqrt{2}%
v_{\phi}\left( \lambda _{5}v_{u}+\lambda _{8}v_{d}\right) \\ 
\sqrt{2}v_{\phi}\left( \lambda _{5}v_{u}+\lambda _{7}v_{d}\right) & \sqrt{2}%
v_{\phi}\left( \lambda _{5}v_{u}+\lambda _{8}v_{d}\right) & 2\lambda
_{6}v_{\phi}^{2} \\ 
&  & 
\end{array}%
\right)  \label{eqn:mass_matrix_CP_even}
\end{equation}
From the mass matrix given above, we find that the CP-even scalar spectrum
is composed of the $125$ GeV\ SM-like Higgs $h^0$ and two non SM CP-even
Higgses $H_{1,2}$.

The CP-odd mass matrix in the basis $\left( \func{Im}%
H_{u}^{0},\func{Im}H_{d}^{0},\func{Im}\Phi \right) $ is given by: 
\begin{equation}
{M}_{CP-\text{odd}}^{2}=\left( 
\begin{array}{ccc}
-\frac{\lambda _{5}v_{d}v_{\phi}^{2}}{2v_{u}} & -\frac{1}{2}\lambda
_{5}v_{\phi}^{2} & -\sqrt{2}\lambda _{5}v_{d}v_{\phi} \\ 
-\frac{1}{2}\lambda _{5}v_{\phi}^{2} & -\frac{\lambda _{5}v_{u}v_{\phi}^{2}}{2v_{d}%
} & -\sqrt{2}\lambda _{5}v_{u}v_{\phi} \\ 
-\sqrt{2}\lambda _{5}v_{d}v_{\phi} & -\sqrt{2}\lambda _{5}v_{u}v_{\phi} & 
-4\lambda _{5}v_{u}v_{d}-4\mu _{sb}^{2}%
\end{array}%
\right)
\end{equation}
In this paper we assume that there is negligible mixing of the $\Phi$ with the two Higgs doublets, so that the physical 
Yukon $\phi$ scalar boson predominantly arises as the real component of the complex singlet field $\Phi$ after it develops its VEV.
	This is a natural assumption in the case that $v_{\phi} \gg v_u,v_d$.
	Alternatively this can be enforced by assuming the coupling terms in the Higgs potential which couple $\Phi$ to the Higgs doublets controlled by $\lambda_5$ to be small.

The charged Higgs mass matrix is given by: 
\begin{equation}
{M}_{\text{charged}}^{2}=\left( 
\begin{array}{cc}
\lambda _4 v_d^2-\frac{\lambda _5 v_d v_{\phi}^2}{2 v_u} & \lambda _4 v_u v_d-%
\frac{1}{2} \lambda _5 v_{\phi}^2 \\ 
\lambda _4 v_u v_d-\frac{1}{2} \lambda _5 v_{\phi}^2 & \lambda _4 v_u^2-\frac{%
\lambda _5 v_u v_{\phi}^2}{2 v_d}%
\end{array}
\right).
\end{equation}

\section{SM Higgs Production Cross-Section}
\label{appendix:h0}

We fully simplify the partial width from The Higgs Hunter's Guide \cite{Gunion:1989we}, taking into account all SM Higgs VEVs and $SU(2)$ gauge coupling factors into a compact expression:
\begin{equation}
	\Gamma(h_0\rightarrow gg) = \frac{\alpha_s^2}{4\pi^3M_{h_0}}\Big|\sum_iy_im_i\big[1+(1-\tau_i)f(\tau_i)\big]\Big|^2\,\,,\,\,\tau_i = \frac{4m_i^2}{M_{h_0}^2}
\end{equation}
$y_i$ represents the coupling at the vertex between the two fermions of flavour $i$ and the SM Higgs boson. The production cross-section from a hard scattering process involving two gluons is then related to the decay width thus;
\begin{equation}
	\frac{d\sigma}{dy}(AB\rightarrow h_0 + X) = \frac{\pi^2\Gamma(h_0\rightarrow gg)}{8M_{h_0}^3}g_A(x_A\,,\,M_{h_0}^2)g_B(x_B\,,\,M_{h_0}^2)
\end{equation}
$g_A$ and $g_B$ are the parton distribution functions (PDFs) of the two gluons, and the fractional momenta $x_A$ and $x_B$ are related to the centre-of-mass energy as per Equation \eqref{eqn:fractional_momenta_definitions}.
\begin{equation}
	x_A = \frac{M_{h_0}e^y}{\sqrt{s}}\quad,\quad x_B = \frac{M_{h_0}e^{-y}}{\sqrt{s}}
	\label{eqn:fractional_momenta_definitions}
\end{equation}
$y$ is the rapidity of the SM Higgs. This gives an expression for the differential cross-section with respect to rapidity:
\begin{equation}
	\frac{d\sigma}{dy}(AB\rightarrow h_0 + X)=\frac{\alpha_s^2}{32\pi M_{h_0}^4}\Big|\sum_iy_im_i\big[1+(1-\tau_i)f(\tau_i)\big]\Big|^2 g_A(x_A\,,\,M_{h_0}^2)g_B(x_B\,,\,M_{h_0}^2).
\end{equation}
In the SM, we do not to make a distinction between the Yukawa coupling $y_i$, and the mass of the top mass $m_t$, since $m_t = y_t v / \sqrt{2}$. However, in the model we scrutinise here, this is no longer true due to mixing between the chiral and VLQs. In general, one defines the Yukawa coupling $y_i$ as the interaction strength between two fermions and the scalar boson. For a generic fermion $f$ coupling to a complex scalar $\phi$, the Yukawa coupling is \footnote{Note given a complex scalar which undergoes Spontaneous Symmetry Breaking (SSB), we would have $\phi = (\varphi + v')/\sqrt{2}$, but in this project we solely write in terms of the complex scalar field and not the real scalar field $\varphi$.} given in Equation \eqref{eqn:def_of_Yukawa_coupling}.
\begin{equation}
	{\cal L} = y_f \bar{f} f \phi
	\label{eqn:def_of_Yukawa_coupling}
\end{equation}
In the VLQ model, one finds a modified top Yukawa coupling to be \cite{Aguilar-Saavedra:2017giu}:
\begin{equation}
	y_t = \frac{\sqrt{2} m_t}{v} \cos ^2 \theta _R ^u = y_t ^0  \cos ^2 \theta _R ^u ,
\end{equation}
which will reduce the cross-section of Higgs production due to the light top state. However, there is also a contribution in the loop due to the heavier VLT, which can be shown to have a Yukawa coupling:
\begin{equation}
	y_T = \frac{\sqrt{2} m_t}{v} \sin ^2 \theta _R ^u .
\end{equation}
In the limit that the VLT has the same mass as SM top, then one would have exact cancellation, and no change to the cross-section. Considering larger VLT masses, there is still some cancellation in this direction. We show the two Feynman diagrams that contribute dominantly to the Higgs production cross-section in our model, in Figure \ref{fig:SM_Higgs_production_gluons}.

In the calculation of the SM Higgs production cross-section in the model under test we include contributions from $t$ and $T$, along with the SM bottom $b$ and the VLB, $B$.

\section{Decay Widths}
\label{sec:appendix_decay_widths}

\subsection{$T$ and $B$ decays}
In this work, for the VLQ decays, we use the results of \cite{Aguilar-Saavedra:2017giu}. We copy their results into the appendix below, and include our additional $Q\rightarrow \phi q$ modes.

Defining $r_x = m_x / m_Q$, where $Q$ is the heavy quark and $x$ one of its decay products, and the function
\begin{equation}
	\lambda(x,y,z) \equiv (x^4 + y^4 + z^4 - 2 x^2 y^2 
	- 2 x^2 z^2 - 2 y^2 z^2) \,,
\end{equation}%
the partial widths for $T$ decays are
\begin{align}
	\Gamma(T \to W^+ b) & = \frac{g^2}{64 \pi}  \frac{M_T}{M_W^2} \lambda(M_T,m_b,M_W)^{1/2} \left\{
	(|V_{Tb}^L|^2+|V_{Tb}^R|^2) \left[ 1+r_W^2-2 r_b^2 -2 r_W^4  + r_b^4 +r_W^2 r_b^2 \right] \right. \nonumber \\ 
	& \left.  -12 r_W^2 r_b \RE V_{Tb}^L V_{Tb}^{R*} \right\}	\,, \nonumber \\ 
	\Gamma(T \to Z t) & = \frac{g^2}{128 \pi c_W^2}  \frac{M_T}{M_Z^2} \lambda(M_T,m_t,M_Z)^{1/2}
	\left\{ (|X_{tT}^L|^2 + |X_{tT}^R|^2) \left[ 1 + r_Z^2 - 2  r_t^2 - 2  r_Z^4  + r_t^4
	+ r_Z^2 r_t^2 \right] \right. \nonumber \\
	& \left.   -12 r_Z^2 r_t \RE X_{tT}^L X_{tT}^{R*}  \right\} \,, \nonumber \\ 
	\Gamma(T \to h^0 t) & = \frac{g^2}{128 \pi}
	\frac{M_T}{M_W^2} \lambda(M_T,m_t,M_{h^0})^{1/2}
	\left\{ (|Y_{tT}^L|^2 + |Y_{tT}^R|^2) \left[1+r_t^2 - r_{h^0}^2 \right] + 4 r_t \RE Y_{tT}^L Y_{tT}^{R*} \right\} \,, \notag \\ 
	\Gamma(T \to \phi t) & = \frac{g^2}{32 \pi}
	\frac{M_T}{M_{Z'}} \lambda(M_T,m_t,M_{\phi})^{1/2}
	\left\{ (|{Y'_{tT}}^L|^2 + |{Y'_{tT}}^R|^2) \left[1+r_t^2 - r_{\phi}^2 \right] + 4 r_t \RE {Y'_{tT}}^L {Y'_{tT}}^{R*} \right\} \,, \notag \\ 
	\label{ec:GammaT}
\end{align}
For the $B$ quark they are analogous,
\begin{align}
	\Gamma(B \to W^- t) & = \frac{g^2}{64 \pi}  \frac{M_B}{M_W^2} \lambda(M_B,m_t,M_W)^{1/2} \left\{
	(|V_{tB}^L|^2+|V_{tB}^R|^2) \left[ 1+r_W^2-2 r_t^2  -2 r_W^4  + r_t^4 +r_W^2 r_t^2
	\right]  \right. \nonumber \\
	&\left. -12 r_W^2 r_t \RE V_{tB}^L V_{tB}^{R*} \right\}
	\,, \nonumber \\
	\Gamma(B \to Z b) & = \frac{g^2}{128 \pi c_W^2}  \frac{M_B}{M_Z^2} \lambda(M_B,m_b,M_Z)^{1/2}
	\left\{ (|X_{bB}^L|^2 + |X_{bB}^R|^2)  \left[ 1 + r_Z^2 - 2  r_b^2 - 2  r_Z^4  + r_b^4
	+ r_Z^2 r_b^2 \right] \right.  \nonumber \\
	&\left. -12 r_Z^2 r_b \RE X_{bB}^L X_{bB}^{R*}  \right\} \,, \nonumber \\
	\Gamma(B \to h^0 b) & = \frac{g^2}{128 \pi}
	\frac{M_B}{M_W^2} \lambda(M_B,m_b,M_{h^0})^{1/2}
	\left\{ (|Y_{bB}^L|^2 + |Y_{bB}^R|^2) \left[1+r_b^2 - r_{h^0}^2 \right] + 4 r_b \RE Y_{bB}^L Y_{bB}^{R*} \right\} \,, \notag \\
	\Gamma(B \to \phi b) & = \frac{g^2}{32 \pi}
	\frac{M_B}{M_{Z'}^2} \lambda(M_B,m_b,M_{\phi})^{1/2}
	\left\{ (|{Y'_{bB}}^L|^2 + |{Y'_{bB}}^R|^2) \left[1+r_b^2 - r_{\phi}^2 \right] + 4 r_b \RE {Y'_{bB}}^L {Y'_{bB}}^{R*} \right\} \,. \notag \\
	\label{ec:GammaB}
\end{align}

\subsection{Yukon decays $\boldmath{\phi} \rightarrow f_1 f_2 $}

Armed with these couplings, we can now consider the partial widths for $\phi$ decay modes, following the process in The Higgs Hunter's Guide \cite{Gunion:1989we} for decay of the SM Higgs boson but adapting for $\phi$. As such, we split possible couplings into their axial and vector pieces in Equation \eqref{eqn:vector_axial_decompotsition}.
\begin{equation}
	V_{\phi f_1 f_2} = i(A + B \gamma ^5)
	\label{eqn:vector_axial_decompotsition}
\end{equation}
The decay of a scalar (of any charge) into two fermions with colour factor $N_c$ goes like Equation \eqref{eqn:partial_width_phi_to_fermions}, where we have labelled the scalar by $\phi$ in anticipation of our application to the $U(1)'$-breaking boson.
\begin{equation}
	\Gamma (\phi \rightarrow f_1 \bar{f_2}) = \frac{ N_c  \lambda\big(m_{f_1},m_{f_2},M_\phi\big)^{1/2}}{4 M_\phi ^3  \pi} \left[ (M_\phi ^2 - m_{f_1} ^2 - m_{f_2} ^2) \frac{A^2 + B^2}{2} - m_{f_1}  m_{f_2} (A^2 - B^2) \right]
	\label{eqn:partial_width_phi_to_fermions}
\end{equation}
Where $\lambda$ is the following kinematic function:
\begin{equation}
	\lambda=\big((m_{f_1}^2 + m_{f_2}^2 - M_\phi^2)^2 - 4m_{f_1}^2 m_{f_2}^2\big).
\end{equation}
This reproduces the usual results for two fermions of the same mass \cite{Gunion:1989we} as seen in Equation \eqref{eqn:scalar_to_same-mass_fermions}.
\begin{equation}
	\Gamma (\phi \rightarrow \bar{f} f) = \frac{N_c g^2 m_f ^2}{32 \pi m_W ^2} \beta ^3 M_\phi = \frac{N_c y_f ^2 }{16 \pi} \beta ^3 M_\phi
	\label{eqn:scalar_to_same-mass_fermions}
\end{equation}
$m_W = gv/2,~m_f=y_f v/\sqrt{2}$. $y_f$ is the usual Yukawa coupling between a scalar and two fermions. The function labelled $\beta$ is purely kinematic and can be expressed as the following:
\begin{equation}
	\beta ^2 = 1 - 4 m_f ^2 / M_\phi ^2 .
\end{equation}

\subsection{Yukon decays $\boldmath{\phi} \rightarrow V_1 V_2 $}
\begin{gather}
	\Gamma(\phi\rightarrow Z'Z') = \frac{g'^2\big(M_\phi^4-4M_\phi^2M_{Z'}^2+12M_{Z'}^4\big)}{32\pi M_{Z'}^2M_\phi}\sqrt{1-\frac{4M_{Z'}^2}{M_\phi^2}},
	\label{eqn:phi_to_Zprimes_partial_width}
\end{gather}
Defining
\begin{equation}
\tau _i = 4 \left( \frac{M_i}{M_\phi} \right) ^2
	\end{equation}
and
\begin{align}
&f(\tau)\equiv
\begin{cases}
	\arcsin^2(\frac{1}{\sqrt{\tau}}), &\mathrm{for}~\tau\geq1\\
	-\frac{1}{4}[\log\frac{1+\sqrt{1-\tau}}{1-\sqrt{1-\tau}}-i\pi]^2, &\mathrm{for}~\tau<1
\end{cases}
\end{align}
then the $\gamma \gamma$ and $gg$ decay modes are \cite{Gunion:1989we}
\begin{align}
	\Gamma\big(\phi\rightarrow\gamma\gamma\big)=\frac{9\alpha_{EM}^2}{8\pi^3}\frac{1}{M_\phi}\Big|\sum_{i}-e_i^2y'_im_i\big[1+\big(1-\tau_i\big)f(\tau_i)\big]\Big|^2 ,
	\label{eqn:phi_to_gamma_gamma_branching_fraction}
\end{align}
\begin{equation}
	\Gamma(\phi\rightarrow gg) = \frac{\alpha_s^2}{4\pi^3M_\phi}\Big|\sum_{i}y'_im_i\big[1+\big(1-\tau_i\big)f(\tau_i)\big]\Big|^2
	\label{eqn:phi_to_gg_branching_fraction}
\end{equation}
Where $e_i$ is the QED charge of the fermion propagating in the loop which couples the scalar to photons and $y_i '$ is defined from
\begin{align}
	\mathcal{L}_\phi\supset-\frac{y'_i}{\sqrt{2}}\overline{\psi}_i\psi_i\phi.
	\label{eqn:Assumed_phi_psi_psi_Lagrangian_coupling}
\end{align}

\subsection{Yukon decay $\mathbf{\phi\rightarrow Z\gamma}$}
We present here the partial width for the Yukon $\phi$ decay to $\gamma Z$ for a VLQ model from \cite{He:2020suf}. The couplings are defined from the following Lagrangian:
\begin{equation}
	\mathcal{L} = -e A_\mu Q_f \bar{f} \gamma^\mu f + e Z_\mu \bar{f} \gamma^\mu (g^f _L P_L + g^f _R) f -\frac{m_t}{{v_\phi}} {\phi} \bar{t} (\kappa _t + i \gamma^5 \tilde{\kappa _t})t+ {\phi} \bar{T} ( y_T + i \gamma^5 \tilde{y} _T)T + {\phi} \bar{t} (y_L ^{tT} P_L + y_R ^{tT} P_R )T
\end{equation}

\begin{align}
	&\Gamma({\phi}\rightarrow\gamma Z)=\frac{G_F\alpha^2m_{\phi}^3}{64\sqrt{2}\pi^3}(1-\frac{m_Z^2}{m_{\phi}^2})^3 |A_t+A_T+A_{tT}|^2.
\end{align}
Where
\begin{align*}
	&\mathcal{A}_t=2N_t^CQ_t(g_{L}^{t}+g_{R}^{t})\kappa_tA_f(\tau_t,\lambda_t)=2N_t^CQ_t\kappa_t\frac{\frac{1}{2}c_L^2+I_3^T(s_L^2+s_R^2)-\frac{4}{3}s_W^2}{s_Wc_W}A_f(\tau_t,\lambda_t),\nonumber\\
	&\mathcal{A}_T=-2N_T^CQ_T\frac{y_{T}{v_\phi}}{M_T}(g_{L}^{T}+g_{R}^{T})A_f(\tau_T,\lambda_T)=-2N_T^CQ_T\frac{y_{T}{v_\phi}}{M_T}\frac{\frac{1}{2}s_L^2+I_3^T(c_L^2+c_R^2)-\frac{4}{3}s_W^2}{s_Wc_W}A_f(\tau_T,\lambda_T),
\end{align*}
\begin{align}
	&\mathcal{A}_{tT}=-4N_T^CQ_T\frac{{v_\phi}}{m_{\phi}^2-m_Z^2}\{m_t\mathrm{Re}(g_{L}^{tT}(y_L^{tT})^{*}+g_{R}^{tT}(y_R^{tT})^{*})[(\frac{m_{\phi}^2-m_Z^2}{2}-m_t^2)C_0(0,m_Z^2,m_{\phi}^2,m_t^2,m_t^2,M_T^2)\nonumber\\
	&-M_T^2C_0(0,m_Z^2,m_{\phi}^2,M_T^2,M_T^2,m_t^2)-m_Z^2\frac{B_0(m_{\phi}^2,m_t^2,M_T^2)-B_0(m_Z^2,m_t^2,M_T^2)}{m_{\phi}^2-m_Z^2}-1]\nonumber\\
	&+M_T\mathrm{Re}(g_{L}^{tT}(y_R^{tT})^{*}+g_{R}^{tT}(y_L^{tT})^{*})[(\frac{m_{\phi}^2-m_Z^2}{2}-M_T^2)C_0(0,m_Z^2,m_{\phi}^2,M_T^2,M_T^2,m_t^2)\nonumber\\
	&-m_t^2C_0(0,m_Z^2,m_{\phi}^2,m_t^2,m_t^2,M_T^2)-m_Z^2\frac{B_0(m_{\phi}^2,m_t^2,M_T^2)-B_0(m_Z^2,m_t^2,M_T^2)}{m_{\phi}^2-m_Z^2}-1]\}.
\end{align}
Here $\tau_i$ and $\lambda_i$ are defined as
\begin{align}
	\tau_f=\frac{4m_f^2}{m_{\phi}^2},\tau_W=\frac{4m_W^2}{m_{\phi}^2},\lambda_f=\frac{4m_f^2}{m_Z^2},\lambda_W=\frac{4m_W^2}{m_Z^2}.
\end{align}
and the $A_f,A_W$ are defined as
\begin{align}
	&A_f(\tau_f,\lambda_f)\equiv I_1(\tau_f,\lambda_f)-I_2(\tau_f,\lambda_f),\nonumber\\
	&I_1(\tau,\lambda)=\frac{\tau\lambda}{2(\tau-\lambda)}+\frac{\tau^2\lambda^2}{2(\tau-\lambda)^2}[f(\tau)-f(\lambda)]+\frac{\tau^2\lambda}{(\tau-\lambda)^2}[g(\tau)-g(\lambda)],\nonumber\\
	&I_2(\tau,\lambda)=-\frac{\tau\lambda}{2(\tau-\lambda)}[f(\tau)-f(\lambda)].
\end{align}
Here $f(\tau)$ is defined as
\begin{align}
	&F_f(\tau_f)\equiv-2\tau_f[1+(1-\tau_f)f(\tau_f)],\quad F_W(\tau_W)\equiv2+3\tau_W+3\tau_W(2-\tau_W)f(\tau_W),\nonumber\\
	&f(\tau)\equiv
	\begin{cases}
		\arcsin^2(\frac{1}{\sqrt{\tau}}), &\mathrm{for}~\tau\geq1\\
		-\frac{1}{4}[\log\frac{1+\sqrt{1-\tau}}{1-\sqrt{1-\tau}}-i\pi]^2, &\mathrm{for}~\tau<1
	\end{cases}.
\end{align} and $g(\tau)$ is defined as
\begin{align}
	&g(\tau)\equiv
	\begin{cases}
		\sqrt{\tau-1}\arcsin(\frac{1}{\sqrt{\tau}}), &\mathrm{for}~\tau\geq1\\
		\frac{1}{2}\sqrt{1-\tau}[\log\frac{1+\sqrt{1-\tau}}{1-\sqrt{1-\tau}}-i\pi], &\mathrm{for}~\tau<1
	\end{cases}.
\end{align}

\begin{align}
	&C_0(0,m_Z^2,m_{\phi}^2,m_t^2,m_t^2,M_T^2)=C_0(0,m_Z^2,m_{\phi}^2,M_T^2,m_t^2,m_t^2)\nonumber\\
	=&-\int_0^1\int_0^1\int_0^1dxdydz\frac{\delta(x+y+z-1)}{[yp_1+z(p_1+p_2)]^2+(x+y)m_t^2+zM_T^2-yp_1^2-z(p_1+p_2)^2}\nonumber\\
	=&-\int_0^1\int_0^1\int_0^1dxdydz\frac{\delta(x+y+z-1)}{yz(m_{\phi}^2-m_Z^2)+z^2m_{\phi}^2+(x+y)m_t^2+zM_T^2-zm_{\phi}^2}.
\end{align}

\begin{figure}[htbp]
	\centering
	\begin{subfigure}{0.24\textwidth}~~~~
		\includegraphics[scale=0.8]{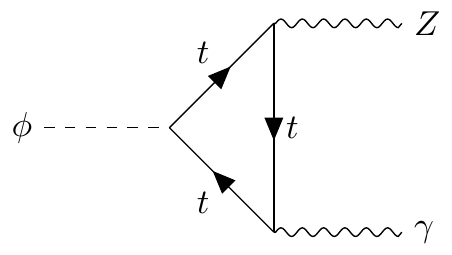}
	\end{subfigure}
	\begin{subfigure}{0.24\textwidth}~~~~
		\includegraphics[scale=0.8]{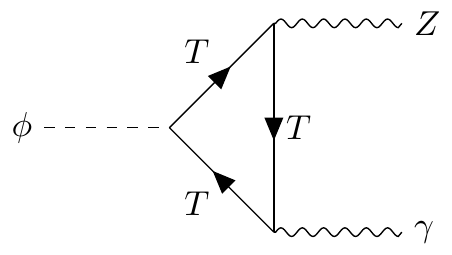}
	\end{subfigure}
	\begin{subfigure}{0.24\textwidth}~~~~
	\includegraphics[scale=0.8]{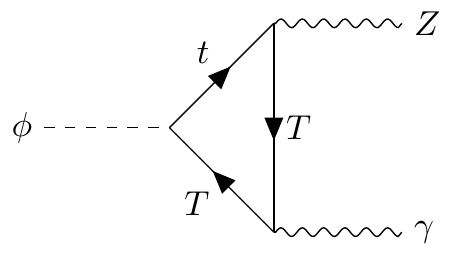}
\end{subfigure}
	\begin{subfigure}{0.24\textwidth}~~~~
	\includegraphics[scale=0.8]{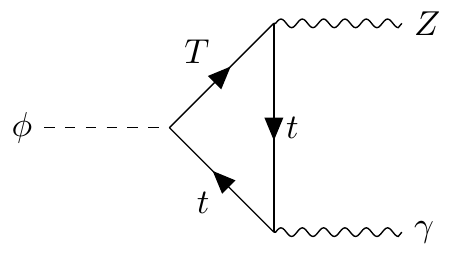}
\end{subfigure}
	\caption{$\phi \rightarrow Z \gamma$ contributions. For the fermion loops, anti-clockwise diagrams should also be included, as well as contributions from the bottom sector by exchanging $t \rightarrow b$ and $T \rightarrow B$. }
	\label{fig:phiZGamma}
\end{figure}

$B_0$ has $\Delta _\epsilon=0$ because $B_0$ occurs in pairs which cancel any finite pieces.


\begin{thebibliography}{99}
 \setlength{\itemsep}{0em}
\bibitem{Dainese:2019rgk}
A.~Dainese, M.~Mangano, A.~B.~Meyer, A.~Nisati, G.~Salam and M.~A.~Vesterinen,
doi:10.23731/CYRM-2019-007

\bibitem{Gianotti:2002xx}
F.~Gianotti, M.~L.~Mangano, T.~Virdee, S.~Abdullin, G.~Azuelos, A.~Ball, D.~Barberis, A.~Belyaev, P.~Bloch and M.~Bosman, \textit{et al.}
Eur. Phys. J. C \textbf{39} (2005), 293-333
doi:10.1140/epjc/s2004-02061-6
[arXiv:hep-ph/0204087 [hep-ph]].

\bibitem{Abada:2019lih}
A.~Abada \textit{et al.} [FCC],
Eur. Phys. J. C \textbf{79} (2019) no.6, 474
doi:10.1140/epjc/s10052-019-6904-3

\bibitem{Chung:2003fi}
D.~J.~H.~Chung, L.~L.~Everett, G.~L.~Kane, S.~F.~King, J.~D.~Lykken and L.~T.~Wang,
Phys. Rept. \textbf{407} (2005), 1-203
doi:10.1016/j.physrep.2004.08.032
[arXiv:hep-ph/0312378 [hep-ph]].

\bibitem{Aaboud:2018pii}
M.~Aaboud \textit{et al.} [ATLAS],
Phys. Rev. Lett. \textbf{121} (2018) no.21, 211801
doi:10.1103/PhysRevLett.121.211801
[arXiv:1808.02343 [hep-ex]].

\bibitem{Romao:2020ocr}
M.~Crispim Rom\~ao, N.~F.~Castro and R.~Pedro,
Eur. Phys. J. C \textbf{81} (2021) no.1, 27
doi:10.1140/epjc/s10052-020-08807-w
[arXiv:2006.05432 [hep-ph]].

\bibitem{Aguilar-Saavedra:2017giu}
J.~A.~Aguilar-Saavedra, D.~E.~L\'opez-Fogliani and C.~Mu\~noz,
JHEP \textbf{06} (2017), 095
doi:10.1007/JHEP06(2017)095
[arXiv:1705.02526 [hep-ph]].

\bibitem{Perelstein:2003wd}
M.~Perelstein, M.~E.~Peskin and A.~Pierce,
Phys. Rev. D \textbf{69} (2004), 075002
doi:10.1103/PhysRevD.69.075002
[arXiv:hep-ph/0310039 [hep-ph]].

\bibitem{Schmaltz:2005ky}
M.~Schmaltz and D.~Tucker-Smith,
Ann. Rev. Nucl. Part. Sci. \textbf{55} (2005), 229-270
doi:10.1146/annurev.nucl.55.090704.151502
[arXiv:hep-ph/0502182 [hep-ph]].

\bibitem{Contino:2006qr}
R.~Contino, L.~Da Rold and A.~Pomarol,
Phys. Rev. D \textbf{75} (2007), 055014
doi:10.1103/PhysRevD.75.055014
[arXiv:hep-ph/0612048 [hep-ph]].

\bibitem{Contino:2006nn}
R.~Contino, T.~Kramer, M.~Son and R.~Sundrum,
JHEP \textbf{05} (2007), 074
doi:10.1088/1126-6708/2007/05/074
[arXiv:hep-ph/0612180 [hep-ph]].

\bibitem{Matsedonskyi:2012ym}
O.~Matsedonskyi, G.~Panico and A.~Wulzer,
JHEP \textbf{01} (2013), 164
doi:10.1007/JHEP01(2013)164
[arXiv:1204.6333 [hep-ph]].

\bibitem{Kaplan:1991dc}
D.~B.~Kaplan,
Nucl. Phys. B \textbf{365} (1991), 259-278
doi:10.1016/S0550-3213(05)80021-5

\bibitem{Martin:2009bg}
S.~P.~Martin,
Phys. Rev. D \textbf{81} (2010), 035004
doi:10.1103/PhysRevD.81.035004
[arXiv:0910.2732 [hep-ph]].

\bibitem{Martin:2010dc}
S.~P.~Martin,
Phys. Rev. D \textbf{82} (2010), 055019
doi:10.1103/PhysRevD.82.055019
[arXiv:1006.4186 [hep-ph]].

\bibitem{Botella:2018aon}
F.~J.~Botella, G.~C.~Branco, M.~Nebot, M.~N.~Rebelo and J.~I.~Silva-Marcos,
[arXiv:1806.02755 [hep-ph]].

\bibitem{Botella:2016ibj}
F.~J.~Botella, G.~C.~Branco, M.~Nebot, M.~N.~Rebelo and J.~I.~Silva-Marcos,
Eur. Phys. J. C \textbf{77} (2017) no.6, 408
doi:10.1140/epjc/s10052-017-4933-3
[arXiv:1610.03018 [hep-ph]].

\bibitem{Ferretti:2006df}
L.~Ferretti, S.~F.~King and A.~Romanino,
JHEP \textbf{11} (2006), 078
doi:10.1088/1126-6708/2006/11/078
[arXiv:hep-ph/0609047 [hep-ph]].

\bibitem{King:2018fcg}
S.~F.~King,
JHEP \textbf{09} (2018), 069
doi:10.1007/JHEP09(2018)069
[arXiv:1806.06780 [hep-ph]].

\bibitem{Aaij:2014ora}
R.~Aaij \textit{et al.} [LHCb],
Phys. Rev. Lett. \textbf{113} (2014), 151601
doi:10.1103/PhysRevLett.113.151601
[arXiv:1406.6482 [hep-ex]].

\bibitem{deAnda:2020hcf}
F.~J.~De Anda and S.~F.~King,
JHEP \textbf{03} (2021), 078
doi:10.1007/JHEP03(2021)078
[arXiv:2010.00100 [hep-ph]].

\bibitem{Hernandez:2021tii}
A.~E.~C.~Hern\'andez, S.~F.~King and H.~Lee,
[arXiv:2101.05819 [hep-ph]].

\bibitem{Tsumura:2009yf}
K.~Tsumura and L.~Velasco-Sevilla,
Phys. Rev. D \textbf{81} (2010), 036012
doi:10.1103/PhysRevD.81.036012
[arXiv:0911.2149 [hep-ph]].

\bibitem{Berger:2014gga}
E.~L.~Berger, S.~B.~Giddings, H.~Wang and H.~Zhang,
Phys. Rev. D \textbf{90} (2014) no.7, 076004
doi:10.1103/PhysRevD.90.076004
[arXiv:1406.6054 [hep-ph]].

\bibitem{Xiao:2014kba}
M.~L.~Xiao and J.~H.~Yu,
Phys. Rev. D \textbf{90} (2014) no.1, 014007
doi:10.1103/PhysRevD.90.014007
[arXiv:1404.0681 [hep-ph]].

\bibitem{Dawson:2012di}
S.~Dawson and E.~Furlan,
Phys. Rev. D \textbf{86} (2012), 015021
doi:10.1103/PhysRevD.86.015021
[arXiv:1205.4733 [hep-ph]].

\bibitem{Aguilar-Saavedra:2013qpa}
J.~A.~Aguilar-Saavedra, R.~Benbrik, S.~Heinemeyer and M.~P\'erez-Victoria,
Phys. Rev. D \textbf{88} (2013) no.9, 094010
doi:10.1103/PhysRevD.88.094010
[arXiv:1306.0572 [hep-ph]].

\bibitem{Spira:1997dg}
M.~Spira,
Fortsch. Phys. \textbf{46} (1998), 203-284
doi:10.1002/(SICI)1521-3978(199804)46:3\ensuremath{<}203::AID-PROP203\ensuremath{>}3.0.CO;2-4
[arXiv:hep-ph/9705337 [hep-ph]].

\bibitem{Aaboud:2018ezd}
M.~Aaboud \textit{et al.} [ATLAS],
Phys. Lett. B \textbf{786} (2018), 114-133
doi:10.1016/j.physletb.2018.09.019
[arXiv:1805.10197 [hep-ex]].

\bibitem{GomezBock:2007hp}
M.~Gomez-Bock, M.~Mondragon, M.~Muhlleitner, M.~Spira and P.~M.~Zerwas,
[arXiv:0712.2419 [hep-ph]].

\bibitem{Coriano:2015sea}
C.~Coriano, L.~Delle Rose and C.~Marzo,
JHEP \textbf{02} (2016), 135
doi:10.1007/JHEP02(2016)135
[arXiv:1510.02379 [hep-ph]].

\bibitem{Accomando:2016sge}
E.~Accomando, C.~Coriano, L.~Delle Rose, J.~Fiaschi, C.~Marzo and S.~Moretti,
JHEP \textbf{07} (2016), 086
doi:10.1007/JHEP07(2016)086
[arXiv:1605.02910 [hep-ph]].

\bibitem{Abreu:1994ria}
P.~Abreu \textit{et al.} [DELPHI],
Z. Phys. C \textbf{65} (1995), 603-618
doi:10.1007/BF01578669

\bibitem{Artuso:2015swg}
M.~Artuso, G.~Borissov and A.~Lenz,
Rev. Mod. Phys. \textbf{88} (2016) no.4, 045002
doi:10.1103/RevModPhys.88.045002
[arXiv:1511.09466 [hep-ph]].

\bibitem{Stelzer:1994ta}
T.~Stelzer and W.~F.~Long,
Comput. Phys. Commun. \textbf{81} (1994), 357-371
doi:10.1016/0010-4655(94)90084-1
[arXiv:hep-ph/9401258 [hep-ph]].

\bibitem{1804.02716}
A.~M.~Sirunyan \textit{et al.} [CMS],
JHEP \textbf{11} (2018), 185
doi:10.1007/JHEP11(2018)185
[arXiv:1804.02716 [hep-ex]].

\bibitem{Catani:2018krb}
S.~Catani, L.~Cieri, D.~de Florian, G.~Ferrera and M.~Grazzini,
JHEP \textbf{04} (2018), 142
doi:10.1007/JHEP04(2018)142
[arXiv:1802.02095 [hep-ph]].

\bibitem{Aad:2019wsl}
G.~Aad \textit{et al.} [ATLAS],
Eur. Phys. J. C \textbf{80} (2020) no.1, 47
doi:10.1140/epjc/s10052-019-7500-2
[arXiv:1909.00761 [hep-ex]].

\bibitem{private_communication}
C.~Shepherd-Themistocleous. Private communication. 2021.



\bibitem{Aaboud:2017jvq}
M.~Aaboud \textit{et al.} [ATLAS],
Phys. Rev. D \textbf{97} (2018) no.7, 072003
doi:10.1103/PhysRevD.97.072003
[arXiv:1712.08891 [hep-ex]].

\bibitem{1804.02610}
A.~M.~Sirunyan \textit{et al.} [CMS],
Phys. Rev. Lett. \textbf{120} (2018) no.23, 231801
doi:10.1103/PhysRevLett.120.231801
[arXiv:1804.02610 [hep-ex]].

\bibitem{Gunion:1989we}
J.~F.~Gunion, H.~E.~Haber, G.~L.~Kane and S.~Dawson,
Front. Phys. \textbf{80} (2000), 1-404
SCIPP-89/13.

\bibitem{He:2020suf}
S.~P.~He,
Phys. Rev. D \textbf{102} (2020) no.7, 075035
doi:10.1103/PhysRevD.102.075035
[arXiv:2004.12155 [hep-ph]].

\end{thebibliography}
\end{document}